\documentclass[final,1p,times]{elsarticle}
\usepackage{amsfonts}
\usepackage{amsmath}
\usepackage{hyperref}
\usepackage{amssymb}
\usepackage[english]{babel}
\usepackage{graphicx}
\usepackage{epsfig}
\usepackage{bm}
\usepackage{longtable}
\usepackage{verbatim}
\usepackage{longtable}

\newcommand{\mathsym}[1]{{}}

\input{tcilatex}
\journal{Nuclear Physics B}

\begin{document}

\begin{frontmatter}



\title{A predictive 3-3-1 model with $A_{4}$ flavor symmetry.}

\author{A. E. C\'arcamo Hern\'andez}
\address{Universidad T\'{e}cnica Federico Santa Mar\'{\i}a\\
and Centro Cient\'{\i}fico-Tecnol\'{o}gico de Valpara\'{\i}so\\
Casilla 110-V, Valpara\'{\i}so, Chile.}
\author{R. Martinez}
\address{Universidad Nacional de Colombia, Departamento de F\'{\i}sica,
Ciudad Universitaria, Bogot\'{a} D.C., Colombia. }
\date{\today }

\begin{abstract}
We propose a predictive model based on the $SU(3)_{C}\otimes
SU(3)_{L}\otimes U(1)_{X}$ gauge group supplemented by the $A_{4}\otimes
Z_{3}\otimes Z_{4}\otimes Z_{6}\otimes Z_{16}$ discrete group, which
successfully describes the SM fermion mass and mixing pattern. The small
active neutrino masses are generated via inverse seesaw mechanism with three
very light Majorana neutrinos. The observed charged fermion mass hierarchy
and quark mixing pattern are originated from the breaking of the $%
Z_{4}\otimes Z_{6}\otimes Z_{16}$ discrete group at very high scale. The
obtained values for the physical observables for both quark and lepton
sectors are in excellent agreement with the experimental data. The model
predicts a vanishing leptonic Dirac CP violating phase as well as an
effective Majorana neutrino mass parameter of neutrinoless double beta
decay, with values $m_{\beta \beta }=$ 2 and 48 meV for the normal and the
inverted neutrino mass hierarchies, respectively.
\end{abstract}


\end{frontmatter}



\section{Introduction}

Despite the great success of the Standard Model (SM), recently confirmed by
the discovery of the $126$ GeV Higgs boson by LHC experiments \cite%
{atlashiggs,cmshiggs,newtevatron,CMS-PAS-HIG-12-020}, there are many aspects
not yet explained such as the origin of the fermion mass and mixing
hierarchy as well as the mechanism responsible for stabilizing the
electroweak scale \cite{SM,PDG}. This discovery of the Higgs scalar field
allows to consider extensions of the SM with additional scalar fields that
can be useful to explain the existence of Dark Matter \cite{BSMtheorieswithDM}.

\quad The Standard Model is a theory with many phenomenological
achievements. However in the Yukawa sector of the SM there are many
parameters related with the fermion masses with no clear dynamical origin.
Because of this reason, it is important to study realistic models that allow
to set up relations among all these parameters of the Yukawa sector.
Discrete flavor symmetries allow to establish ansatz that explain the flavor
problem, for recent reviews see Refs. \cite%
{King:2013eh,Altarelli:2010gt,Ishimori:2010au}. These discrete flavor
symmetries may be crucial in building models of fermion mixing that address
the flavor problem. Non abelian discrete flavor symmetries arise in string
theories due to the discrete features of the fixed points of the orbifolds 
\cite{StringsandDS}. For instance, the discrete $D_{4}$ group is originated
in the $S^{1}/Z_{2}$ orbifold \cite{StringsandDS}.

\quad Besides that, another of the greatest mysteries in particle physics is
the existence of three fermion families at low energies. The quark mixing
angles are small whereas the leptonic mixing angles are large. Models based
on the gauge symmetry $SU(3)_{C}\otimes SU(3)_{L}\otimes U(1)_{X}$ have the
feature of being vectorlike with three families of fermions and are
therefore anomaly free \cite{331-pisano,331-frampton,331-long,M-O,anomalias}%
. When the electric charge is defined in the linear combination of the $%
SU(3)_{L}$ generators $T_{3}$ and $T_{8}$, it is a free parameter,
independent of the anomalies ($\beta $). The choice of this parameter
defines the charge of the exotic particles. Choosing $\beta =-\frac{1}{\sqrt{%
3}}$, the third component of the weak lepton triplet is a neutral field $\nu
_{R}^{C}$, which allows to build the Dirac matrix with the usual field $\nu
_{L}$ of the weak doublet. If one introduces a sterile neutrino $N_{R}$ in
the model, then it is possible to generate light neutrino masses via inverse
seesaw mechanism. The 3-3-1 models with $\beta =-\frac{1}{\sqrt{3}}$ have
the advantange of providing an alternative framework to generate neutrino
masses, where the neutrino spectrum includes the light active sub-eV scale
neutrinos as well as sterile neutrinos which could be dark matter candidates
if they are light enough or candidates for detection at the LHC, if their
masses are at the TeV scale. This interesting feature makes the 3-3-1 models
very interesting, since if the TeV scale sterile neutrinos are found at the
LHC, these models can be very strong candidates for unraveling the mechanism
responsible for electroweak symmetry breaking. Furthermore, the 3-3-1 models can provide an explanation for the $750$ GeV diphoton excess recently reported by ATLAS and CMS \cite{diphotonexcess331} as well as for the $2$ TeV diboson excess found by ATLAS \cite{dibosonexcess331}.

\quad Neutrino oscillation experiments \cite{PDG,An:2012eh,Abe:2011sj,Adamson:2011qu,Abe:2011fz,Ahn:2012nd} indicate
that there are at least two massive active neutrinos and at most one
massless active neutrino. In the mass eigenstates, it is necessary for the
solar neutrinos oscillations that $\delta
m_{sun}^{2}=m_{21}^{2}=m_{2}^{2}-m_{1}^{2}$ where $m_{2}^{2}-m_{1}^{2}>0$.
For the atmospheric neutrinos oscillations it is required that $\delta
m_{atm}^{2}=m_{31}^{2}=m_{3}^{2}-m_{1}^{2}$ where the difference can be
possitive (normal hierarchy) or negative (inverted hierarchy). Neutrino
oscillations do not give information neither on the absolute value of the
neutrino mass nor on the Majorana or Dirac nature of the neutrino. However
there are neutrino mass bounds arising from cosmology \cite{Ade:2013zuv},
tritium beta decay \cite{Kraus:2004zw} and double beta decay \cite%
{Auger:2012ar,Abt:2004yk,Ackermann:2012xja,Alessandria:2011rc,KamLANDZen:2012aa,Auty:2013:zz,Guiseppe:2011me,Bilenky:2014uka,Albert:2014fya}%
.

\quad The neutrino masses and mixings are known from neutrino oscillations,
which depend on the squared neutrino mass diferences and not on the absolute
value of the neutrino masses. The global fits of the available data from the
Daya Bay \cite{An:2012eh}, T2K \cite{Abe:2011sj}, MINOS \cite{Adamson:2011qu}%
, Double CHOOZ \cite{Abe:2011fz} and RENO \cite{Ahn:2012nd} neutrino
oscillation experiments, constrain the neutrino mass squared splittings and
mixing parameters \cite{Forero:2014bxa}. The current neutrino data on
neutrino mixing parameters can be very well accommodated in the approximated
tribimaximal mixing matrix, 
\begin{equation}
U_{TBM}=\left( 
\begin{array}{ccc}
\sqrt{\frac{2}{3}} & \frac{1}{\sqrt{3}} & 0 \\ 
-\frac{1}{\sqrt{6}} & \frac{1}{\sqrt{3}} & -\frac{1}{\sqrt{2}} \\ 
-\frac{1}{\sqrt{6}} & \frac{1}{\sqrt{3}} & \frac{1}{\sqrt{2}}%
\end{array}%
\right),\label{TBM-ansatz}
\end{equation}
which is consistent with two large mixing angles and one very small one of order zero. Specifically, the mixing angles predicted by the
tribimaximal mixing matrix satisfy $\left( \sin ^{2}\theta _{12}\right)
_{TBM}=\frac{1}{3}$, $\left( \sin ^{2}\theta _{23}\right) _{TBM}=\frac{1}{2}$%
, and $\left( \sin ^{2}\theta _{13}\right) _{TBM}=0$. However, the 3-3-1 model
is not able to generate the tribimaximal matrix structure. Because of this
reason, discrete symmetry groups \cite{discrete-lepton,6HDMA4,SU5A4,331S3,discrete-quark,s3pheno,Delta27,T7,SU5T7,Tprime}
that act on the fermion families are imposed with the aim to generate ansatz
that reproduce these matrices. One of the most promising discrete flavor
groups is $A_{4}$, since it is the smallest symmetry with one
three-dimensional and three distinct one-dimensional irreducible
representations, where the three families of fermions can be accommodated
rather naturally. Another approach to describe the fermion mass and mixing
pattern consists in postulating particular mass matrix textures (see Ref 
\cite{textures} for some works considering textures). Besides that, models
with Multi-Higgs sectors, Grand Unification, Extradimensions and
Superstrings as well as with horizontal symmetries have been proposed in the
literature \cite{King:2013eh,GUT,Extradim,String,horizontal} to explain the
observed pattern of fermion masses and mixings. 

\quad In this paper we propose a version of the $SU(3)_{C}\times
SU(3)_{L}\times U(1)_{X}$ model with an additional flavor symmetry group %
\mbox{$A_{4}\otimes Z_3\otimes Z_4\otimes Z_6\otimes Z_{16}$} and an
extended scalar sector needed in order to reproduce the specific patterns of
mass matrices in the fermion sector that successfully account for fermion
masses and mixings. The particular role of each additional scalar field and
the corresponding particle assignments under the symmetry group of the model
are explained in detail in Sec. \ref{model}. The model we are building with
the aforementioned discrete symmetries, preserves the content of particles
of the minimal 3-3-1 model, but we add additional very heavy scalar fields
with quantum numbers that allow to build Yukawa terms invariant under the
local and discrete groups. This generates the predictive and viable textures
that explain the 18 physical observables in the quark and lepton sectors,
i.e., the 9 charged fermion masses, 2 neutrino mass squared splittings, 3
lepton mixing parameters, 3 quark mixing angles and 1 CP violating phase of
the CKM quark mixing matrix. Our model successfully describes the prevailing
pattern of the SM fermion masses and mixing. 

\quad The content of this paper is organized as follows. In Sec. \ref{model}
we outline the proposed model. In Sec. \ref{leptonmassesandmixing} we
discusss lepton masses and mixings and show our corresponding results. Our
results for the masses and mixings in the quark sector followed by a
numerical analysis are presented in Sec. \ref{quarkmassesandmixing}. Finally in Sec. \ref{conclusions}, we state our conclusions. In \ref{ap1} we present a brief description of the $A_{4}$ group.

\section{The model}

\label{model} We extend the $SU(3)_{C}\otimes SU(3)_{L}\otimes U(1)_{X}$
group of the minimal 3-3-1 model by adding an extra flavor symmetry group %
\mbox{$A_{4}\otimes Z_3\otimes Z_4\otimes Z_6\otimes Z_{16}$}, in such a way
that the full symmetry $\mathcal{G}$ experiences a three-step spontaneous
breaking, as follows: 
\begin{eqnarray}
&&\mathcal{G}=SU(3)_{C}\otimes SU\left( 3\right) _{L}\otimes U\left(
1\right) _{X}\otimes A_{4}\otimes Z_{3}\otimes Z_{4}\otimes Z_{6}\otimes
Z_{16}  \label{Group} \\
&&{\xrightarrow{\Lambda _{int}}}SU(3)_{C}\otimes SU\left( 3\right)
_{L}\otimes U\left( 1\right) _{X}\otimes Z_{3}{\xrightarrow{v_{\chi }}}%
SU(3)_{C}\otimes SU\left( 2\right) _{L}\otimes U\left( 1\right) _{Y}  \notag
\\
&&{\xrightarrow{v_{\eta },v_{\rho }}}SU(3)_{C}\otimes U\left( 1\right) _{Q},
\notag
\end{eqnarray}%
where the different symmetry breaking scales satisfy the following hierarchy 
$v_{\eta },v_{\rho }\ll v_{\chi }\ll \Lambda _{int}.$ 

\quad We define the electric charge in our 3-3-1 model in terms of the $SU(3)$
generators and the identity, as follows: 
\begin{equation}
Q=T_{3}-\frac{1}{\sqrt{3}}T_{8}+XI,
\end{equation}%
with $I=Diag(1,1,1)$, $T_{3}=\frac{1}{2}Diag(1,-1,0)$ and $T_{8}=(\frac{1}{2%
\sqrt{3}})Diag(1,1,-2)$.

The anomaly cancellation of $SU(3)_{L}$ requires that the two families of
quarks be accommodated in $3^{\ast }$ irreducible representations (irreps).
Besides that, the number of $3^{\ast }$ irreducible representations is six,
as follows from the quark colors. We accommodate the other family of quarks
into a $3$ irreducible representation. Furthermore, we have six $3$ irreps
taking into account the three families of leptons. Thus, the $SU(3)_{L}$
representations are vector like and free of anomalies. Having anomaly free $%
U(1)_{X}$ representations requires that the quantum numbers for the fermion
families be assigned in such a way that the combination of the $U(1)_{X}$
representations with other gauge sectors cancels anomalies. Consequently, to
avoid anomalies, the fermions have to be accommodated into the following $(SU(3)_{C},SU(3)_{L},U(1)_{X})$ left- and right-handed representations: 
\begin{align}
Q_{L}^{1,2}& =%
\begin{pmatrix}
D^{1,2} \\ 
-U^{1,2} \\ 
J^{1,2} \\ 
\end{pmatrix}%
_{L}:(3,3^{\ast },0),\hspace{0.5cm}Q_{L}^{3}=%
\begin{pmatrix}
U^{3} \\ 
D^{3} \\ 
T \\ 
\end{pmatrix}%
_{L}:(3,3,1/3),\hspace{0.5cm}L_{L}^{1,2,3}=%
\begin{pmatrix}
\nu ^{1,2,3} \\ 
e^{1,2,3} \\ 
(\nu ^{1,2,3})^{c} \\ 
\end{pmatrix}%
_{L}:(1,3,-1/3),  \notag \\
& 
\begin{array}{c}
D_{R}^{1,2}:(3,1,-1/3), \\ 
U_{R}^{1,2}:(3,1,2/3), \\ 
J_{R}^{1,2}:(3,1,-1/3), \\ 
\end{array}%
\hspace{0.5cm}%
\begin{array}{c}
U_{R}^{3}:(3,1,2/3), \\ 
D_{R}^{3}:(3,1,-1/3), \\ 
T_{R}:(3,1,2/3), \\ 
\end{array}%
\hspace{0.5cm}  \notag \\
& 
\begin{array}{c}
e_{R}:(1,1,-1), \\ 
N_{R}^{1}:(1,1,0), \\ 
\end{array}%
\hspace{0.5cm}%
\begin{array}{c}
\mu _{R}:(1,1,-1), \\ 
N_{R}^{2}:(1,1,0), \\ 
\end{array}%
\hspace{0.5cm}%
\begin{array}{c}
\tau _{R}:(1,1,-1), \\ 
N_{R}^{3}:(1,1,0), \\ 
\end{array}%
\end{align}%
where $U_{L}^{i}$ and $D_{L}^{i}$ for $i=1,2,3$ are three up- and down-type
quark components in the flavor basis, while $\nu _{L}^{i}$ and $e_{L}^{i}$ ($%
e_{L},\mu _{L},\tau _{L}$) are the neutral and charged lepton families. The
right-handed fermions are assigned as $SU(3)_{L}$ singlets representations
having $U(1)_{X}$ quantum numbers equal to their electric charges.
Furthermore, the fermion spectrum of the model includes as heavy fermions: a
single flavor quark $T$ with electric charge $2/3$, two flavor quarks $%
J^{1,2}$ with charge $-1/3$, three neutral Majorana leptons $(\nu
^{1,2,3})_{L}^{c}$ and three right-handed Majorana leptons $N_{R}^{1,2,3}$
(see Ref. \cite{catano} for a recent discussion about neutrino masses via
double and inverse see-saw mechanism for a 3-3-1 model).

\quad The 3-3-1 models extend the scalar sector of the SM into three $3$'s
irreps of $SU(3)_{L}$, where one heavy triplet $\chi $ acquires a vaccuum
expectation value (VEV) at the TeV scale, $v_{\chi }$, breaking the $SU(3)_{L}\times U(1)_{X}$ symmetry down to the $SU(2)_{L}\times U(1)_{Y}$
electroweak group of the SM and then giving masses to the non SM fermions
and gauge bosons; and two lighter triplet fields $\eta $ and $\rho $ that get
VEVs $v_{\eta }$ and $v_{\rho }$, respectively, at the electroweak scale
thus generating the mass for the fermion and gauge sector of the SM. We
enlarge the scalar sector of the minimal 3-3-1 model by introducing 14 $SU(3)_{L}$ scalar singlets, namely, $\xi _{j}$, $\zeta _{j}$, $S_{j}$ , $\varphi $,$\ \Delta $,$\ \phi $,$\ \tau $ and $\sigma $ ($j=1,2,3$). 

\quad The scalars of our model are accommodated into the following $%
[SU(3)_{L},U(1)_{X}]$ representations: 
\begin{align}
\chi & =%
\begin{pmatrix}
\chi _{1}^{0} \\ 
\chi _{2}^{-} \\ 
\frac{1}{\sqrt{2}}(\upsilon _{\chi }+\xi _{\chi }\pm i\zeta _{\chi }) \\ 
\end{pmatrix}%
:(3,-1/3),\hspace{0.7cm}\xi _{j}:(1,0),\hspace{0.7cm}\tau :(1,0),\hspace{%
0.7cm}\varphi :(1,0),\hspace{0.7cm}j=1,2,3,  \notag \\
\rho & =%
\begin{pmatrix}
\rho _{1}^{+} \\ 
\frac{1}{\sqrt{2}}(\upsilon _{\rho }+\xi _{\rho }\pm i\zeta _{\rho }) \\ 
\rho _{3}^{+} \\ 
\end{pmatrix}%
:(3,2/3),\hspace{0.7cm}\zeta _{j}:(1,0),\hspace{0.7cm}\phi :(1,0),\hspace{%
0.7cm}\Delta :(1,0),\hspace{0.7cm}j=1,2,3,  \notag \\
\eta & =%
\begin{pmatrix}
\frac{1}{\sqrt{2}}(\upsilon _{\eta }+\xi _{\eta }\pm i\zeta _{\eta }) \\ 
\eta _{2}^{-} \\ 
\eta _{3}^{0}%
\end{pmatrix}%
:(3,-1/3),\hspace{1cm}S_{j}:(1,0),\hspace{1cm}\sigma \sim (1,0),\hspace{1cm}%
j=1,2,3.  \label{331-scalar}
\end{align}%
\quad The scalar fields are grouped into triplet and singlet representions
of $A_{4}$. The scalar fields of our model have the following assignments
under $A_{4}\otimes Z_{3}\otimes Z_{4}\otimes Z_{6}\otimes Z_{16}$: 
\begin{eqnarray}
\eta &\sim &\left( \mathbf{1,}e^{-\frac{2i\pi }{3}},1,1,1\right) ,\hspace{%
0.5cm}\rho \sim \left( \mathbf{1,}e^{\frac{2i\pi }{3}},1,1,1\right) ,\hspace{%
0.5cm}\chi \sim \left( \mathbf{1},1,1,1,1\right) ,\hspace{1cm}  \notag \\
\xi &\sim &\left( \mathbf{3},1,1,1,-1\right) ,\hspace{0.5cm}\zeta \sim
\left( \mathbf{3,}1,1,1,e^{\frac{i\pi }{8}}\right) ,\hspace{0.5cm}S\sim
\left( \mathbf{3,}e^{-\frac{2i\pi }{3}},1,1,e^{\frac{i\pi }{8}}\right) ,%
\hspace{0.5cm}\sigma \sim \left( \mathbf{1,}1,1,1,e^{-\frac{i\pi }{8}%
}\right) ,  \notag \\
\varphi &\sim &\left( \mathbf{1,}1,1,e^{-\frac{i\pi }{3}},1\right) ,\hspace{%
0.5cm}\Delta \sim \left( \mathbf{1,}1,-1,e^{-\frac{i\pi }{3}},1\right) ,%
\hspace{0.5cm}\phi \sim \left( \mathbf{1}^{\prime }\mathbf{,}1,i,1,1\right) ,%
\hspace{0.5cm}\tau \sim \left( \mathbf{1}^{\prime \prime }\mathbf{,}%
1,i,1,1\right) ,  \label{scalarassignments}
\end{eqnarray}%
where the numbers in boldface are dimensions of the $A_{4}$ irreducible
representations.

The leptons transform under $A_{4}\otimes Z_{3}\otimes Z_{4}\otimes
Z_{6}\otimes Z_{16}$ as: 
\begin{eqnarray}
L_{L} &\sim &\left( \mathbf{3,}e^{\frac{2i\pi }{3}},1,1,-1\right) ,\hspace{%
1cm}e_{R}\sim \left( \mathbf{1},1,1,1,e^{\frac{7i\pi }{8}}\right) ,\hspace{%
1cm}\mu _{R}\sim \left( \mathbf{1}^{\prime },1,1,1,i\right) ,  \notag \\
\tau _{R} &\sim &\left( \mathbf{1}^{\prime \prime }\mathbf{,}1,1,1,e^{\frac{%
i\pi }{4}}\right) ,\hspace{1cm}N_{R}\sim \left( \mathbf{3,}e^{\frac{2i\pi }{3%
}},1,1,-1\right) .  \label{leptonassignments}
\end{eqnarray}

Note that left handed leptons are unified into a $A_{4}$ triplet
representation $\mathbf{3}$, whereas the right handed charged leptons are
assigned into different $A_{4}$ singlets, i.e, $\mathbf{1}$, $\mathbf{1}%
^{\prime }$ and $\mathbf{1}^{\prime \prime }$. Furthermore, the right handed
Majorana neutrinos are unified into a $A_{4}$ triplet representation. 

The $A_{4}\otimes Z_{3}\otimes Z_{4}\otimes Z_{6}\otimes Z_{16}$ assignments
for the quark sector are: 
\begin{eqnarray}
Q_{L}^{1} &\sim &\left( \mathbf{1,}1,1,1,e^{-\frac{i\pi }{8}}\right) ,%
\hspace{1cm}Q_{L}^{2}\sim \left( \mathbf{1}^{\prime }\mathbf{,}%
1,1,1,1\right) ,\hspace{1cm}Q_{L}^{3}\sim \left( \mathbf{1}^{\prime \prime }%
\mathbf{,}1,1,1,1\right) ,  \notag \\
U_{R}^{1} &\sim &\left( \mathbf{1,}e^{-\frac{2\pi i}{3}},1,1,e^{\frac{7i\pi 
}{8}}\right) ,\hspace{1cm}U_{R}^{2}\sim \left( \mathbf{1}^{\prime }\mathbf{,}%
e^{-\frac{2\pi i}{3}},1,1,i\right) ,\hspace{1cm}U_{R}^{3}\sim \left( \mathbf{%
1}^{\prime \prime }\mathbf{,}e^{-\frac{2\pi i}{3}},1,1,1\right) ,  \notag \\
D_{R}^{1} &\sim &\left( \mathbf{1,}e^{\frac{2\pi i}{3}},1,-1,e^{\frac{i\pi }{%
8}}\right) ,\hspace{1cm}D_{R}^{2}\sim \left( \mathbf{1,}e^{\frac{2\pi i}{3}%
},1,-1,1\right) ,\hspace{1cm}D_{R}^{3}\sim \left( \mathbf{1}^{\prime \prime }%
\mathbf{,}e^{\frac{2\pi i}{3}},1,-1,1\right)  \notag \\
T_{R} &\sim &\left( \mathbf{1}^{\prime \prime }\mathbf{,}1,1,1,1\right) ,%
\hspace{1cm}J_{R}^{1}\sim \left( \mathbf{1}^{\prime }\mathbf{,}%
1,1,1,1\right) \hspace{1cm}J_{R}^{2}\sim \left( \mathbf{1}^{\prime \prime }%
\mathbf{,}1,1,1,1\right) .  \label{Quarkassignments}
\end{eqnarray}%
With the above particle content, the following relevant Yukawa terms for the
quark and lepton sector arise: 
\begin{eqnarray}
-\mathcal{L}_{Y}^{\left( Q\right) } &=&y_{11}^{\left( U\right) }\overline{Q}%
_{L}^{1}\rho ^{\ast }U_{R}^{1}\frac{\sigma ^{8}}{\Lambda ^{8}}%
+y_{22}^{\left( U\right) }\overline{Q}_{L}^{2}\rho ^{\ast }U_{R}^{2}\frac{%
\sigma ^{4}}{\Lambda ^{2}}+y_{33}^{\left( U\right) }\overline{Q}_{L}^{3}\eta
U_{R}^{3}  \notag \\
&&+y_{11}^{\left( D\right) }\overline{Q}_{L}^{1}\eta ^{\ast }D_{R}^{1}\frac{%
\tau \phi \Delta ^{3}\sigma ^{2}}{\Lambda ^{7}}+y_{12}^{\left( D\right) }%
\overline{Q}_{L}^{1}\eta ^{\ast }D_{R}^{2}\frac{\tau \phi \Delta ^{3}\sigma 
}{\Lambda ^{6}}+y_{13}^{\left( D\right) }\overline{Q}_{L}^{1}\eta ^{\ast
}D_{R}^{3}\frac{\phi ^{2}\Delta ^{3}\sigma }{\Lambda ^{6}}  \notag \\
&&+y_{21}^{\left( D\right) }\overline{Q}_{L}^{2}\eta ^{\ast }D_{R}^{1}\frac{%
\tau ^{2}\Delta ^{3}\sigma }{\Lambda ^{6}}+y_{22}^{\left( D\right) }%
\overline{Q}_{L}^{2}\eta ^{\ast }D_{R}^{2}\frac{\tau ^{2}\Delta ^{3}}{%
\Lambda ^{5}}+y_{23}^{\left( D\right) }\overline{Q}_{L}^{2}\eta ^{\ast
}D_{R}^{3}\frac{\phi ^{2}\Delta ^{3}}{\Lambda ^{5}}  \notag \\
&&+y_{31}^{\left( D\right) }\overline{Q}_{L}^{3}\rho D_{R}^{1}\frac{\phi
^{2}\Delta ^{3}\sigma }{\Lambda ^{6}}+y_{32}^{\left( D\right) }\overline{Q}%
_{L}^{3}\rho D_{R}^{2}\frac{\phi ^{2}\Delta ^{3}}{\Lambda ^{5}}%
+y_{33}^{\left( D\right) }\overline{Q}_{L}^{3}\rho D_{R}^{3}\frac{\varphi
^{3}}{\Lambda ^{3}}  \notag \\
&&+y^{\left( T\right) }\overline{Q}_{L}^{3}\chi T_{R}+y_{1}^{\left( J\right)
}\overline{Q}_{L}^{1}\chi ^{\ast }J_{R}^{1}+y_{2}^{\left( J\right) }%
\overline{Q}_{L}^{2}\chi ^{\ast }J_{R}^{2}  \label{Lyquarks}
\end{eqnarray}%
\begin{eqnarray}
-\mathcal{L}_{Y}^{\left( L\right) } &=&h_{\rho e}^{\left( L\right) }\left( 
\overline{L}_{L}\rho \xi \right) _{\mathbf{\mathbf{1}}}e_{R}\frac{\sigma ^{7}%
}{\Lambda ^{8}}+h_{\rho \mu }^{\left( L\right) }\left( \overline{L}_{L}\rho
\xi \right) _{\mathbf{1}^{\prime \prime }}\mu _{R}\frac{\sigma ^{4}}{\Lambda
^{5}}+h_{\rho \tau }^{\left( L\right) }\left( \overline{L}_{L}\rho \xi
\right) _{\mathbf{1^{\prime }}}\tau _{R}\frac{\sigma ^{2}}{\Lambda ^{3}} 
\notag \\
&+&h_{\chi }^{\left( L\right) }\left( \overline{L}_{L}\chi N_{R}\right) _{%
\mathbf{1}}+\frac{1}{2}h_{1N}\left( \overline{N}_{R}N_{R}^{C}\right) _{%
\mathbf{\mathbf{1}}}\frac{\left( \eta ^{\dagger }\cdot \eta ^{\ast }\right)
+x\left( \rho ^{T}\cdot \rho \right) }{\Lambda }+h_{2N}\left( \overline{N}%
_{R}N_{R}^{C}\right) _{\mathbf{3s}}\frac{S\sigma }{\Lambda }  \notag \\
&+&h_{\rho }\varepsilon _{abc}\left( \overline{L}_{L}^{a}\left(
L_{L}^{C}\right) ^{b}\right) _{\mathbf{3s}}\left( \rho ^{\ast }\right) ^{c}%
\frac{\zeta \sigma }{\Lambda ^{2}}+H.c,  \label{Lylepton}
\end{eqnarray}%
where the dimensionless couplings $y_{ii}^{\left( U\right) }$, $%
y_{ij}^{\left( D\right) }$ ($i,j=1,2,3$), $y^{\left( T\right) }$, $%
y_{1}^{\left( J\right) }$, $y_{2}^{\left( J\right) }$, $h_{\rho e}^{\left(
L\right) }$, $h_{\rho \mu }^{\left( L\right) }$, $h_{\rho \tau }^{\left(
L\right) }$, $h_{\chi }^{\left( L\right) }$, $h_{1N}$, $x$, $h_{2N}$ and $h_{\rho
}$ are $\mathcal{O}(1)$ parameters. Here we assumed that all Yukawa
couplings are real, excepting $y_{13}^{\left( D\right) }$, $y_{31}^{\left(
D\right) }$ and $h_{\rho \tau }^{\left( L\right) }$ which are assumed to be
complex.

\quad Although the flavor discrete groups in Eq. (\ref{Group}) look rather
sofisticated, each discrete group factor plays its own role in generating
predictive fermion textures that successfully account for the low energy
fermion flavor data. To describe the pattern of fermion masses and mixing
angles, one needs to postulate particular Yukawa textures. As we will see in
the next sections, the predictive textures for the lepton and quark sectors
will give rise to the experimentally observed deviation of the tribimaximal
mixing pattern and to quark mixing emerging only from the down type quark
sector, respectively. A candidate for generating specific Yukawa textures is
the $A_{4}$ flavor symmetry, which needs to be supplemented by the $%
Z_{3}\otimes Z_{4}\otimes Z_{6}\otimes Z_{16}$ discrete group. As we will
see in the next sections, this predictive setup can successfully
account for fermion masses and mixings. The inclusion of the $A_{4}$
discrete group reduces the number of parameters in the Yukawa and scalar
sector of the $SU(3)_{C}\otimes SU(3)_{L}\otimes U(1)_{X}$ model making it
more predictive. We choose $A_{4}$ since it is the smallest discrete group
with a three-dimensional irreducible representation and 3 distinct
one-dimensional irreducible representations, which allows to naturally
accommodate the three fermion families. We unify the left-handed leptons in
the $A_{4}$ triplet representation and the right-handed leptons are assigned
to $A_{4}$ singlets. Regarding the quark sector, we assign quarks into $%
A_{4} $ singlet representations. In what follows we describe the role of
each discrete cyclic group factor introduced in our model. 
The $Z_{3}$ symmetry separates the $A_{4}$ scalar triplets participating in
the Yukawa interactions for charged leptons from those ones participating in
the neutrino Yukawa interactions. Besides that, the $Z_{3}$ symmetry avoids
mixings between SM quarks and exotic quarks since the right handed exotic
quarks are neutral under this symmetry whereas the right handed SM quarks
have non trivial $Z_{3}$ charges. Thus the $Z_{3}$ symmetry decouples the SM
quarks from the exotic quarks resulting in a reduction of quark sector model
parameters. Furthermore, the $Z_{4}$ symmetry is also important for reducing
the number of quark sector model parameters, since due to this symmetry, the 
$SU(3) _{L}$ scalar singlets $A_{4}$ nontrivial singlets only
appear in the down type quark Yukawa terms. Consequently this $Z_{4}$
symmetry together with the $A_{4}$ assignments for quarks described in Eq. (\ref{Quarkassignments}), results in a diagonal up type quark mass matrix,
thus giving rise to a quark mixing only emerging from the down type quark
sector. The $Z_{6}$ symmetry is crucial for explaining the hierarchy between
the SM down and SM up type quarks without tuning the SM down type quark
Yukawa couplings, since it is the smallest cyclic symmetry that allows $%
\frac{\varphi ^{3}}{\Lambda ^{3}}$ in the Yukawa term that generates the
bottom quark mass, which is $\lambda ^{3}\frac{v}{\sqrt{2}}$ ($\lambda
=0.225$ is one of the Wolfenstein parameters) times a $\mathcal{O}(1)$
parameter. The $Z_{16}$ symmetry gives rise to the observed hierarchy among
charged fermion masses and quark mixing angles. It is worth mentioning that
the properties of the $Z_{N}$ groups imply that the $Z_{16}$ symmetry is the
smallest cyclic symmetry that allows to build the Yukawa term $\overline{Q}%
_{L}^{1}\rho ^{\ast }U_{R}^{1}\frac{\sigma ^{8}}{\Lambda ^{8}}$ of dimension
twelve from a $\frac{\sigma ^{8}}{\Lambda ^{8}}$ insertion on the $\overline{%
Q}_{L}^{1}\rho ^{\ast }U_{R}^{1}$ operator, crucial to get the required $%
\lambda ^{8}$ suppression (where $\lambda =0.225$ is one of the Wolfenstein
parameters) needed to naturally explain the smallness of the up quark mass.
Regarding the charged lepton sector, let us note that the five dimensional
Yukawa operators $\frac{1}{\Lambda }\left( \overline{L}_{L}\rho \xi \right)
_{\mathbf{\mathbf{1}}}e_{R}$, $\frac{1}{\Lambda }\left( \overline{L}_{L}\rho
\xi \right) _{\mathbf{1}^{\prime \prime }}\mu _{R}$ and $\frac{1}{\Lambda }%
\left( \overline{L}_{L}\rho \xi \right) _{\mathbf{1^{\prime }}}\tau _{R}$
are $A_{4}$ invariant but do not preserve the $Z_{16}$ symmetry, as follows
from the charges assignments given by Eqs. (\ref{scalarassignments}) and (%
\ref{leptonassignments}).

\quad In what follows we comment about the possible VEVs patterns for the $A_{4}$
scalar triplets $\xi $, $\zeta $\ and $S$. Here we assume a hierarchy
between the VEVs of the $A_{4}$ scalar triplets $\xi $, $\zeta $\ and $S$,
i.e., $v_{S}<<v_{\zeta }<<v_{\xi }$ , which implies that the mixing angles
of these scalar triplets are very small since they are suppressed by the
ratios of their VEVs, which is a consequence of the method of recursive
expansion proposed in Ref. \cite{grimus}. Consequently, we can neglect the
mixing between the $A_{4}$ scalar triplets $\xi $, $\zeta $ and $S$, and
treat their corresponding scalar potentials independently. The relevant
terms determining the VEV directions of any $A_{4}$ acalar triplet are: 
\begin{eqnarray}
V\left( \Sigma \right) &=&-\mu _{\Sigma }^{2}\left( \Sigma \Sigma ^{\ast
}\right) _{\mathbf{1}}+\kappa _{\Sigma ,1}\left( \Sigma \Sigma ^{\ast
}\right) _{\mathbf{1}}\left( \Sigma \Sigma ^{\ast }\right) _{\mathbf{1}%
}+\kappa _{\Sigma ,2}\left( \Sigma \Sigma \right) _{\mathbf{1}}\left( \Sigma
^{\ast }\Sigma ^{\ast }\right) _{\mathbf{1}}+\kappa _{\Sigma ,3}\left(
\Sigma \Sigma ^{\ast }\right) _{\mathbf{1}^{\prime }}\left( \Sigma \Sigma
^{\ast }\right) _{\mathbf{1}^{\prime \prime }}  \notag \\
&&+\kappa _{\Sigma ,4}\left[ \left( \Sigma \Sigma \right) _{\mathbf{1}%
^{\prime }}\left( \Sigma ^{\ast }\Sigma ^{\ast }\right) _{\mathbf{1}^{\prime
\prime }}+h.c\right] +\kappa _{\Sigma ,5}\left[ \left( \Sigma \Sigma \right)
_{\mathbf{1}^{\prime \prime }}\left( \Sigma ^{\ast }\Sigma ^{\ast }\right) _{%
\mathbf{1}^{\prime }}+h.c\right]  \notag \\
&&+\kappa _{\Sigma ,6}\left( \Sigma \Sigma ^{\ast }\right) _{\mathbf{3s}%
}\left( \Sigma \Sigma ^{\ast }\right) _{\mathbf{3s}}+\kappa _{\Sigma
,7}\left( \Sigma \Sigma \right) _{\mathbf{3s}}\left( \Sigma ^{\ast }\Sigma
^{\ast }\right) _{\mathbf{3s}}.  \label{ScalarpotentialA4triplet}
\end{eqnarray}
where $\Sigma =\xi $, $\zeta $, $S.$

The part of the scalar potential for each $A_{4}$ scalar triplet has 8 free
parameters: 1 bilinear and 7 quartic couplings. The minimization conditions
of the scalar potential for a $A_{4}$ triplet yield the following relations:
\begin{eqnarray}
\frac{\partial \left\langle V\left( \Sigma \right) \right\rangle }{\partial 
\text{$v_{\Sigma _{1}}$}} &=&-2v_{\Sigma _{1}}\mu _{\Sigma }^{2}+4\kappa
_{\Sigma ,1}\text{$v_{\Sigma _{1}}$}\left( \text{$v_{\Sigma
_{1}}^{2}+v_{\Sigma _{2}}^{2}+v_{\Sigma _{3}}^{2}$}\right) +2\kappa _{\Sigma
,3}\text{$v_{\Sigma _{1}}$}\left( 2\text{$v_{\Sigma _{1}}^{2}-v_{\Sigma
_{2}}^{2}-v_{\Sigma _{3}}^{2}$}\right)   \notag \\
&&+4\kappa _{\Sigma ,2}\text{$v_{\Sigma _{1}}$}\left[ \text{$v_{\Sigma
_{1}}^{2}+v_{\Sigma _{2}}^{2}\cos \left( 2\theta _{\Sigma _{1}}-2\theta
_{\Sigma _{2}}\right) +v_{\Sigma _{3}}^{2}\cos \left( 2\theta _{\Sigma
_{1}}-2\theta _{\Sigma _{3}}\right) $}\right] +8\kappa _{\Sigma ,7}\text{$%
v_{\Sigma _{1}}$}\left( v_{\Sigma _{2}}^{2}+v_{\Sigma _{3}}^{2}\right)  
\notag \\
&&+4\left( \kappa _{\Sigma ,4}+\kappa _{\Sigma ,5}\right) \text{$v_{\Sigma
_{1}}$}\left[ 2\text{$v_{\Sigma _{1}}^{2}-v_{\Sigma _{2}}^{2}\cos \left(
2\theta _{\Sigma _{1}}-2\theta _{\Sigma _{2}}\right) -v_{\Sigma
_{3}}^{2}\cos \left( 2\theta _{\Sigma _{1}}-2\theta _{\Sigma _{3}}\right) $}%
\right]   \notag \\
&&+4\kappa _{\Sigma ,6}\text{$v_{\Sigma _{1}}$}\left[ \text{$v_{\Sigma
_{2}}^{2}\left\{ 1+\cos \left( 2\theta _{\Sigma _{1}}-2\theta _{\Sigma
_{2}}\right) \right\} +v_{\Sigma _{3}}^{2}\left\{ 1+\text{$\cos \left(
2\theta _{\Sigma _{1}}-2\theta _{\Sigma _{3}}\right) $}\right\} $}\right]  
\notag \\
&=&0,  \notag \\
\frac{\partial \left\langle V\left( \Sigma \right) \right\rangle }{\partial 
\text{$v_{\Sigma _{2}}$}} &=&-2v_{\Sigma _{2}}\mu _{\Sigma }^{2}+4\kappa
_{\Sigma ,1}\text{$v_{\Sigma _{2}}$}\left( \text{$v_{\Sigma
_{1}}^{2}+v_{\Sigma _{2}}^{2}+v_{\Sigma _{3}}^{2}$}\right) +2\kappa _{\Sigma
,3}\text{$v_{\Sigma _{2}}$}\left( 2\text{$v_{\Sigma _{2}}^{2}-v_{\Sigma
_{1}}^{2}-v_{\Sigma _{3}}^{2}$}\right)   \notag \\
&&+4\kappa _{\Sigma ,2}\text{$v_{\Sigma _{2}}$}\left[ v_{\Sigma _{2}}^{2}+%
\text{$v_{\Sigma _{1}}^{2}\cos \left( 2\theta _{\Sigma _{2}}-2\theta
_{\Sigma _{1}}\right) +v_{\Sigma _{3}}^{2}\cos \left( 2\theta _{\Sigma
_{2}}-2\theta _{\Sigma _{3}}\right) $}\right] +8\kappa _{\Sigma ,7}\text{$%
v_{\Sigma _{2}}$}\left( v_{\Sigma _{1}}^{2}+v_{\Sigma _{3}}^{2}\right)  
\notag \\
&&+4\left( \kappa _{\Sigma ,4}+\kappa _{\Sigma ,5}\right) \text{$v_{\Sigma
_{2}}$}\left[ 2v_{\Sigma _{2}}^{2}-\text{$v_{\Sigma _{1}}^{2}\cos \left(
2\theta _{\Sigma _{2}}-2\theta _{\Sigma _{1}}\right) -v_{\Sigma
_{3}}^{2}\cos \left( 2\theta _{\Sigma _{2}}-2\theta _{\Sigma _{3}}\right) $}%
\right]   \notag \\
&&+4\kappa _{\Sigma ,6}\text{$v_{\Sigma _{2}}$}\left[ \text{$v_{\Sigma
_{1}}^{2}\left\{ 1+\cos \left( 2\theta _{\Sigma _{2}}-2\theta _{\Sigma
_{1}}\right) \right\} +v_{\Sigma _{3}}^{2}\left\{ 1+\text{$\cos \left(
2\theta _{\Sigma _{2}}-2\theta _{\Sigma _{3}}\right) $}\right\} $}\right]  
\notag \\
&=&0,  \notag \\
\frac{\partial \left\langle V\left( \Sigma \right) \right\rangle }{\partial 
\text{$v_{\Sigma _{3}}$}} &=&-2v_{\Sigma _{3}}\mu _{\Sigma }^{2}+4\kappa
_{\Sigma ,1}\text{$v_{\Sigma _{3}}$}\left( \text{$v_{\Sigma
_{1}}^{2}+v_{\Sigma _{2}}^{2}+v_{\Sigma _{3}}^{2}$}\right) +2\kappa _{\Sigma
,3}\text{$v_{\Sigma _{3}}$}\left( 2\text{$v_{\Sigma _{3}}^{2}-v_{\Sigma
_{1}}^{2}-v_{\Sigma _{2}}^{2}$}\right)   \notag \\
&&+4\kappa _{\Sigma ,2}\text{$v_{\Sigma _{3}}$}\left[ v_{\Sigma _{2}}^{2}+%
\text{$v_{\Sigma _{1}}^{2}\cos \left( 2\theta _{\Sigma _{1}}-2\theta
_{\Sigma _{2}}\right) +v_{\Sigma _{2}}^{2}\cos \left( 2\theta _{\Sigma
_{3}}-2\theta _{\Sigma _{2}}\right) $}\right] +8\kappa _{\Sigma ,7}\text{$%
v_{\Sigma _{3}}$}\left( v_{\Sigma _{1}}^{2}+v_{\Sigma _{2}}^{2}\right)  
\notag \\
&&\text{$+$}4\left( \kappa _{\Sigma ,4}+\kappa _{\Sigma ,5}\right) \text{$%
v_{\Sigma _{3}}$}\left[ 2v_{\Sigma _{3}}^{2}-\text{$v_{\Sigma _{1}}^{2}\cos
\left( 2\theta _{\Sigma _{1}}-2\theta _{\Sigma _{2}}\right) -v_{\Sigma
_{2}}^{2}\cos \left( 2\theta _{\Sigma _{3}}-2\theta _{\Sigma _{2}}\right) $}%
\right]   \notag \\
&&+4\kappa _{\Sigma ,6}\text{$v_{\Sigma _{3}}$}\left[ \text{$v_{\Sigma
_{1}}^{2}\left\{ 1+\cos \left( 2\theta _{\Sigma _{1}}-2\theta _{\Sigma
_{2}}\right) \right\} +v_{\Sigma _{2}}^{2}\left\{ 1+\text{$\cos \left(
2\theta _{\Sigma _{3}}-2\theta _{\Sigma _{2}}\right) $}\right\} $}\right]  
\notag \\
&=&0.  \label{DV}
\end{eqnarray}
where $\left\langle \Sigma \right\rangle =\left( \text{$v_{\Sigma
_{1}}e^{i\theta _{\Sigma _{1}}},v_{\Sigma _{2}}e^{i\theta _{\Sigma
_{2}}},v_{\Sigma _{3}}e^{i\theta _{\Sigma _{3}}}$}\right)$. 
Here in order to simplify the analysis, we restrict to the simplest case of
zero phases in the VEV patterns of the $A_{4}$ triplet scalars, i.e., $\theta _{\Sigma _{1}}=\theta _{\Sigma _{2}}=\theta _{\Sigma _{3}}=0$. Then,
from the scalar potential minimization equations given by Eq. (\ref{DV}),
the following relations are obtained:
\begin{eqnarray}
\left[ 3\kappa _{\Sigma ,3}-4\left( \kappa _{\Sigma ,6}+\kappa _{\Sigma
,7}\right) +6\left( \kappa _{\Sigma ,4}+\kappa _{\Sigma ,5}\right) \right]
\left( \text{$v_{\Sigma _{1}}^{2}-v_{\Sigma _{2}}^{2}$}\right)  &=&0,  \notag
\\
\left[ 3\kappa _{\Sigma ,3}-4\left( \kappa _{\Sigma ,6}+\kappa _{\Sigma
,7}\right) +6\left( \kappa _{\Sigma ,4}+\kappa _{\Sigma ,5}\right) \right]
\left( \text{$v_{\Sigma _{1}}^{2}-v_{\Sigma _{3}}^{2}$}\right)  &=&0,  \notag
\\
\left[ 3\kappa _{\Sigma ,3}-4\left( \kappa _{\Sigma ,6}+\kappa _{\Sigma
,7}\right) +6\left( \kappa _{\Sigma ,4}+\kappa _{\Sigma ,5}\right) \right]
\left( \text{$v_{\Sigma _{2}}^{2}-v_{\Sigma _{3}}^{2}$}\right)  &=&0.
\label{DVb}
\end{eqnarray}
From the relations given by Eq. (\ref{DVb}) and setting $\kappa _{\zeta
,3}=\frac{4}{3}\left( \kappa _{\zeta ,6}+\kappa _{\zeta ,7}\right) -2\left(
\kappa _{\zeta ,4}+\kappa _{\zeta ,5}\right) $, we obtain the following VEV
pattern: 
\begin{equation}
\left\langle \xi \right\rangle =\frac{v_{\xi }}{\sqrt{3}}\left( 1,1,1\right)
,\hspace{1cm}\left\langle \zeta \right\rangle =\frac{v_{\zeta }}{\sqrt{2}}%
\left( 1,0,1\right) ,\hspace{1cm}\left\langle S\right\rangle =\frac{v_{S}}{%
\sqrt{3}}\left( 1,1,-1\right) .  \label{VEV}
\end{equation}%
%
%
%
%
%
%
%
%
%
%
%
%
%
%
%
%
%
%
%
%
%
%
%
%
%
%
%
%
%
%
In the case of 
$\xi $, this is a vacuum configuration preserving a $Z_{3}$ subgroup of $%
A_{4}$, which has been extensively studied in many $A_{4}$ flavor models
(for recent reviews see Refs. \cite%
{King:2013eh,Altarelli:2010gt,Ishimori:2010au})%
. The VEV pattern for the $A_{4}$ triplet scalar $\zeta $ is similar to the
one previously studied in an $A_{4}$ and $T_{7}$ flavor $SU(5)$ GUT models \cite{SU5A4,SU5T7}
and in a 6HDM with $A_{4}$ flavor symmetry \cite{6HDMA4}. As we
will see in the next section, the VEV patterns for the $A_{4}$ triplets $\xi 
$, $\zeta $ and $S$ given in Eq. \ref{VEV}) are crucial to get a predictive
model that successfully reproduces the experimental values of the physical
observables in the lepton sector. 

\quad Furthermore we assume that these $SU(3)_{L}$ scalar singlets get VEVs
at a scale $\Lambda _{int}$ much larger than $v_{\chi }$ (which is of the
order of the TeV scale), with the exception of $S_{j}$ ($j=1,2,3$), which
get VEVs much smaller than the electroweak symmetry breaking scale $v=246$
GeV. The VEVs of the $\xi _{j}$ ($j=1,2,3$), $\varphi $, $\Delta $, $\phi $, 
$\tau $ and $\sigma $ scalar singlets break the $SU(3)_{C}\otimes
SU(3)_{L}\otimes U(1)_{X}\otimes A_{4}\otimes Z_{3}\otimes Z_{4}\otimes
Z_{6}\otimes Z_{16}$ symmetry down to $SU(3)_{C}\otimes SU(3)_{L}\otimes
U(1)_{X}\otimes Z_{3}$ at the scale $\Lambda _{int}$.

From the expressions given above, and using the vacuum configuration for the 
$A_{4}$ scalar triplets given in Eq. (\ref{VEV}), we find the following
relations:
\begin{eqnarray}
\mu _{\xi }^{2} &=&\frac{2}{3}\left[ 3\left( \kappa _{\xi ,1}+\kappa _{\xi
,2}\right) +4\left( \kappa _{\xi ,6}+\kappa _{\xi ,7}\right) \right] v_{\xi
}^{2},  \notag \\
\mu _{\zeta }^{2} &=&\frac{2}{3}\left[ 3\left( \kappa _{\zeta ,1}+\kappa
_{\zeta ,2}\right) +4\left( \kappa _{\zeta ,6}+\kappa _{\zeta ,7}\right) %
\right] v_{\zeta }^{2},  \notag \\
\mu _{S}^{2} &=&\frac{2}{3}\left[ 3\left( \kappa _{S,1}+\kappa _{S,2}\right)
+4\left( \kappa _{S,6}+\kappa _{S,7}\right) \right] v_{S}^{2}.
\label{musquad}
\end{eqnarray}

These results show that the VEV directions for the three $A_{4}$ triplets,
i.e., $\xi $, $\zeta $ and $S$ scalars in Eq. (\ref{VEV}), are consistent
with a global minimum of the scalar potential (\ref{ScalarpotentialA4triplet}%
) of our model for a large region of parameter space.

\quad Besides that, as the hierarchy among charged fermion masses and quark mixing angles emerges from the breaking of the $Z_{4}\otimes
Z_{6}\otimes Z_{16}$ discrete group, we set the VEVs of the $SU(3)_{L}$
singlet scalar fields $\xi $,$\ \varphi $,$\ \Delta $,$\ \phi $,$\ \tau $
and $\sigma $, with respect to the Wolfenstein parameter $\lambda =0.225$
and the model cutoff $\Lambda $, as follows: 
\begin{equation}
v_{\varphi }\sim v_{\tau }\sim v_{\phi }\sim v_{\Delta }\sim v_{\xi }\sim
v_{\sigma }\sim \Lambda _{int}=\lambda \Lambda .  \label{VEVsinglets}
\end{equation}%
Furthermore, we assume that the $A_{4}$ scalar triplets participating in the
neutrino Yukawa interactions have VEVs much smaller than the electroweak
symmetry breaking scale. Besides that, as previously mentioned, we assume a
hierarchy among the VEVs of the two $A_{4}$ scalar triplets participating in
the neutrino Yukawa terms. Consequently, as we will see in the next section,
the Majorana neutrinos acquire very small masses and thus an inverse seesaw
mechanism for the generation of light active neutrino masses, takes place.
Therefore, we have the following hierarchy among the VEVs of the scalar
fields in our model: 
\begin{equation}
v_{S}<<v_{\zeta }<<v_{\rho }\sim v_{\eta }\sim v<<v_{\chi }<<\Lambda _{int}.
\end{equation}

In what follows, we briefly comment about the low energy scalar sector of
our model. The renormalizable low energy scalar potential of the model is
given by:%
%
%
%
%
%
%
%
%
%
\begin{eqnarray}
V_{H}&=&\mu _{\chi }^{2}(\chi ^{\dagger }\chi )+\mu _{\eta }^{2}(\eta
^{\dagger }\eta )+\mu _{\rho }^{2}(\rho ^{\dagger }\rho )+f\left( \chi
_{i}\eta _{j}\rho _{k}\varepsilon ^{ijk}+H.c.\right) +\lambda _{1}(\chi
^{\dagger }\chi )(\chi ^{\dagger }\chi )  \notag \\
&&+\lambda _{2}(\rho ^{\dagger }\rho )(\rho ^{\dagger }\rho )+\lambda
_{3}(\eta ^{\dagger }\eta )(\eta ^{\dagger }\eta )+\lambda _{4}(\chi
^{\dagger }\chi )(\rho ^{\dagger }\rho )+\lambda _{5}(\chi ^{\dagger }\chi
)(\eta ^{\dagger }\eta )  \notag \\
&&+\lambda _{6}(\rho ^{\dagger }\rho )(\eta ^{\dagger }\eta )+\lambda
_{7}(\chi ^{\dagger }\eta )(\eta ^{\dagger }\chi )+\lambda _{8}(\chi
^{\dagger }\rho )(\rho ^{\dagger }\chi )+\lambda _{9}(\rho ^{\dagger }\eta
)(\eta ^{\dagger }\rho ).  \label{v00}
\end{eqnarray}

After the symmetry breaking takes place, it is found that the scalar mass
eigenstates are related with the weak scalar states by: \cite{331-long, M-O}%
: 
\begin{eqnarray}
\begin{pmatrix}
G_{1}^{\pm } \\ 
H_{1}^{\pm } \\ 
\end{pmatrix}%
=R_{\beta _{T}}%
\begin{pmatrix}
\rho _{1}^{\pm } \\ 
\eta _{2}^{\pm } \\ 
\end{pmatrix}
&,&\hspace{0.3cm}%
\begin{pmatrix}
G_{1}^{0} \\ 
A_{1}^{0} \\ 
\end{pmatrix}%
=R_{\beta _{T}}%
\begin{pmatrix}
\zeta _{\rho } \\ 
\zeta _{\eta } \\ 
\end{pmatrix}%
,\hspace{0.3cm}%
\begin{pmatrix}
H_{1}^{0} \\ 
h^{0} \\ 
\end{pmatrix}%
=R_{\alpha _{T}}%
\begin{pmatrix}
\xi _{\rho } \\ 
\xi _{\eta } \\ 
\end{pmatrix}%
,  \label{331-mass-scalar-a} \\
\begin{pmatrix}
G_{2}^{0} \\ 
H_{2}^{0} \\ 
\end{pmatrix}%
=R%
\begin{pmatrix}
\chi _{1}^{0} \\ 
\eta _{3}^{0} \\ 
\end{pmatrix}
&,&\hspace{0.3cm}%
\begin{pmatrix}
G_{3}^{0} \\ 
H_{3}^{0} \\ 
\end{pmatrix}%
=R%
\begin{pmatrix}
\zeta _{\chi } \\ 
\xi _{\chi } \\ 
\end{pmatrix}%
,\hspace{0.3cm}%
\begin{pmatrix}
G_{2}^{\pm } \\ 
H_{2}^{\pm } \\ 
\end{pmatrix}%
=R%
\begin{pmatrix}
\chi _{2}^{\pm } \\ 
\rho _{3}^{\pm } \\ 
\end{pmatrix}%
,  \label{331-mass-scalar-b}
\end{eqnarray}

with

\begin{equation}
R_{\alpha _{T}(\beta _{T})}=\left( 
\begin{array}{cc}
\cos \alpha _{T}(\beta _{T}) & \sin \alpha _{T}(\beta _{T}) \\ 
-\sin \alpha _{T}(\beta _{T}) & \cos \alpha _{T}(\beta _{T})%
\end{array}%
\right) ,\hspace{2cm}R=\left( 
\begin{array}{cc}
-1 & 0 \\ 
0 & 1%
\end{array}%
\right) ,
\end{equation}
where $\tan \beta _{T}=v_{\eta }/v_{\rho }$, and $\tan 2\alpha
_{T}=M_1/(M_2-M_3)$ with: 
\begin{eqnarray}
M_1&=&4\lambda _6 v_{\eta}v_{\rho}+2\sqrt{2}fv_{\chi},  \notag \\
M_2&=&4\lambda _2 v_{\rho}^2-\sqrt{2}fv_{\chi}\tan \beta _T ,  \notag \\
M_3&=&4\lambda _3 v_{\eta}^2-\sqrt{2}fv_{\chi}/\tan \beta _T .
\end{eqnarray}

It is noteworthy to mention that the our model has the following scalar
states at low energies: 4 massive charged Higgs ($H_{1}^{\pm }$, $H_{2}^{\pm
}$), one CP-odd Higgs ($A_{1}^{0}$), 3 neutral CP-even Higgs ($%
h^{0},H_{1}^{0},H_{3}^{0}$) and 2 neutral Higgs ($H_{2}^{0},\overline{H}%
_{2}^{0}$) bosons. We identify the scalar $h^{0}$ with the SM-like $126$ GeV
Higgs boson discovered at the LHC. Let us note that the neutral Goldstone
bosons $G_{1}^{0}$, $G_{3}^{0}$, $G_{2}^{0}$, $\overline{G}_{2}^{0}$
correspond to the longitudinal components of the $Z$, $Z^{\prime }$, $K^{0}$
and $\overline{K}^{0}$gauge bosons, respectively. Besides that, the charged
Goldstone bosons $G_{1}^{\pm }$ and $G_{2}^{\pm }$ are associated to the
longitudinal components of the $W^{\pm }$ and $K^{\pm }$ gauge bosons,
respectively \cite{331-pisano,M-O}.

%
%
%
%
%
%
%
%
%
%
%

\newpage
\section{Lepton masses and mixings}

\label{leptonmassesandmixing} From Eq. (\ref{Lylepton}) and taking into
account that the VEV pattern of the $A_4$ triplet, $SU(3)_{L}$ singlet scalar field $\xi$ satisfies Eq. (\ref{VEV}) with the nonvanishing VEVs of the $SU(3)_{L}$ singlet scalars $\xi$ and $\sigma$, set to be equal to $\lambda \Lambda $ (being $\Lambda $ the cutoff of our model) as indicated by
Eq. (\ref{VEVsinglets}), we find that the charged lepton mass matrix is
given by: 
\begin{equation}
M_{l}=R_{lL}^{\dag }P_{l}diag\left( m_{e},m_{\mu },m_{\tau }\right) ,\hspace{%
0.4cm}R_{lL}=\frac{1}{\sqrt{3}}\left( 
\begin{array}{ccc}
1 & 1 & 1 \\ 
1 & \omega & \omega ^{2} \\ 
1 & \omega ^{2} & \omega%
\end{array}%
\right) ,\hspace{0.4cm}\omega =e^{\frac{2\pi i}{3}},\hspace{0.4cm}%
P_{l}=\left( 
\begin{array}{ccc}
1 & 0 & 0 \\ 
0 & 1 & 0 \\ 
0 & 0 & e^{i\alpha }%
\end{array}%
\right) ,  \label{Ml}
\end{equation}%
being $\alpha$ the complex phase of $h_{\rho \tau }^{\left( L\right) }$, and the
charged lepton masses are: 
\begin{equation}
m_{e}=a_{1}^{\left( l\right) }\lambda ^{8}\frac{v}{\sqrt{2}},\hspace{1cm}%
m_{\mu }=a_{2}^{\left( l\right) }\lambda ^{5}\frac{v}{\sqrt{2}},\hspace{1cm}%
m_{\tau }=a_{3}^{\left( l\right) }\lambda ^{3}\frac{v}{\sqrt{2}}.
\label{leptonmasses}
\end{equation}

where $\lambda =0.225$ is one of the Wolfenstein parameters, $v=246$ GeV the
scale of electroweak symmetry breaking and $a_{i}^{\left( l\right) }$ ($%
i=1,2,3$) are $\mathcal{O}(1)$ parameters. Let us note that the charged
lepton masses are linked with the scale of electroweak symmetry breaking
through their power dependence on the Wolfenstein parameter $\lambda =0.225$%
, with $\mathcal{O}(1)$ coefficients. Furthermore, from the lepton Yukawa
terms given in Eq. (\ref{Lylepton}) it is easy to see that our model does
not feature flavor changing leptonic neutral Higgs decays. Consequently, our
model cannot explain the recently reported anomaly in the $h\rightarrow \mu
\tau $ decay, implying that a measurement of its branching fraction will be
decisive for its exclusion.%

\quad Regarding the neutrino sector, we can write the neutrino mass terms
as: 
\begin{equation}
-\mathcal{L}_{mass}^{\left( \nu \right) }=\frac{1}{2}\left( 
\begin{array}{ccc}
\overline{\nu _{L}^{C}} & \overline{\nu _{R}} & \overline{N_{R}}%
\end{array}%
\right) M_{\nu }\left( 
\begin{array}{c}
\nu _{L} \\ 
\nu _{R}^{C} \\ 
N_{R}^{C}%
\end{array}%
\right) +H.c,  \label{Lnu}
\end{equation}%
where the neutrino mass matrix is constrained from the $A_{4}$ flavor
symmetry and has the following form: 
\begin{equation}
M_{\nu }=\left( 
\begin{array}{ccc}
0_{3\times 3} & M_{D} & 0_{3\times 3} \\ 
M_{D}^{T} & 0_{3\times 3} & M_{\chi } \\ 
0_{3\times 3} & M_{\chi }^{T} & M_{R}%
\end{array}%
\right) ,
\end{equation}%
and the submatrices are given by: 
\begin{eqnarray}
M_{D} &=&\frac{h_{\rho }v_{\rho }v_{\zeta }v_{\sigma }}{2\Lambda ^{2}}\left( 
\begin{array}{ccc}
0 & 1 & 0 \\ 
-1 & 0 & -1 \\ 
0 & 1 & 0%
\end{array}%
\right) ,\hspace{0.5cm}M_{R}=\left( 
\begin{array}{ccc}
h_{1N}\frac{v_{\eta }^{2}+xv_{\rho }^{2}}{\Lambda ^{2}}v_{\sigma } & -h_{2N}%
\frac{v_{S}v_{\sigma }}{\sqrt{3\Lambda }} & h_{2N}\frac{v_{S}v_{\sigma }}{%
\sqrt{3}\Lambda } \\ 
-h_{2N}\frac{v_{S}v_{\sigma }}{\sqrt{3\Lambda }} & h_{1N}\frac{v_{\eta
}^{2}+xv_{\rho }^{2}}{\Lambda ^{2}}v_{\sigma } & h_{2N}\frac{v_{S}v_{\sigma }%
}{\sqrt{3}\Lambda } \\ 
h_{2N}\frac{v_{S}v_{\sigma }}{\sqrt{3}\Lambda } & h_{2N}\frac{v_{S}v_{\sigma
}}{\sqrt{3}\Lambda } & h_{1N}\frac{v_{\eta }^{2}+xv_{\rho }^{2}}{\Lambda ^{2}%
}v_{\sigma }%
\end{array}%
\right),\notag \\
M_{\chi } &=&h_{\chi }^{\left( L\right) }\frac{v_{\chi }}{\sqrt{2}}\left( 
\begin{array}{ccc}
1 & 0 & 0 \\ 
0 & 1 & 0 \\ 
0 & 0 & 1%
\end{array}%
\right) .
\end{eqnarray}%
As previously mentioned, we assume in our model that the $SU(3)_{L}$ scalar
singlet, $A_{4}$ triplet $S$ interacting with the right handed Majorana
neutrinos gets a very small vacuum expectation value, much smaller than the
electroweak symmetry breaking scale, which results in very small masses for
these Majorana neutrinos. Consequently, this setup can generate small active
neutrino masses through an inverse seesaw mechanism.

\quad As shown in detail in Ref. \cite{catano}, the full rotation matrix is
approximately given by: 
\begin{equation}
\mathbb{U}=%
\begin{pmatrix}
R_{\nu } & B_{3}U_{\chi } & B_{2}U_{R} \\ 
-\frac{(B_{2}^{\dagger }+B_{3}^{\dagger })}{\sqrt{2}}R_{\nu } & \frac{(1-S)}{%
\sqrt{2}}U_{\chi } & \frac{(1+S)}{\sqrt{2}}U_{R} \\ 
-\frac{(B_{2}^{\dagger }-B_{3}^{\dagger })}{\sqrt{2}}R_{\nu } & \frac{(-1-S)%
}{\sqrt{2}}U_{\chi } & \frac{(1-S)}{\sqrt{2}}U_{R}%
\end{pmatrix}%
,  \label{U}
\end{equation}%
where 
\begin{equation}
S=-\frac{1}{2\sqrt{2}h_{\chi }^{\left( L\right) }v_{\chi }}M_{R},\hspace{1cm}%
\hspace{1cm}B_{2}\simeq B_{3}\simeq \frac{1}{h_{\chi }^{\left( L\right)
}v_{\chi }}M_{D}^{\ast },
\end{equation}%
and the physical neutrino mass matrices are: 
\begin{eqnarray}
M_{\nu }^{\left( 1\right) } &=&M_{D}\left( M_{\chi }^{T}\right)
^{-1}M_{R}M_{\chi }^{-1}M_{D}^{T},  \label{Mnu1} \\
M_{\nu }^{\left( 2\right) } &=&-M_{\chi }^{T}+\frac{1}{2}M_{R},\hspace{1cm}%
\hspace{1cm}M_{\nu }^{\left( 3\right) }=M_{\chi }^{T}+\frac{1}{2}M_{R},
\label{Mnu2}
\end{eqnarray}%
where $M_{\nu }^{\left( 1\right) }$ corresponds to the active neutrino mass
matrix whereas $M_{\nu }^{\left( 2\right) }$ and $M_{\nu }^{\left( 3\right)
} $ are the exotic Dirac neutrino mass matrices. Note that the physical
eigenstates include three active neutrinos and six exotic neutrinos. The
exotic neutrinos are pseudo-Dirac, with masses $\sim \pm M_{\chi }^{T}$ and
a small splitting $M_{R}$. Furthermore, $R_{\nu }$, $U_{R}$ and $U_{\chi }$
are the rotation matrices which diagonalize $M_{\nu }^{\left( 1\right) }$, $%
M_{\nu }^{\left( 2\right) }$ and $M_{\nu }^{\left( 3\right) }$, respectively 
\cite{catano}.

\quad From Eq. (\ref{Mnu1}) it follows that the light active neutrino mass
matrix is given by: 
\begin{eqnarray}
M_{\nu }^{\left( 1\right) } &=&-\frac{h_{\rho }^{2}v_{\rho }^{2}v_{\zeta
}^{2}v_{\sigma }}{2h_{\chi }^{\left( L\right) }v_{\chi }^{2}\Lambda ^{3}}%
\left( 
\begin{array}{ccc}
h_{1N}\frac{v_{\chi }^{2}}{\Lambda } & 0 & h_{1N}\frac{v_{\chi }^{2}}{%
\Lambda } \\ 
0 & h_{1N}\frac{v_{\chi }^{2}}{\Lambda }+\frac{2h_{2N}}{\sqrt{3}}v_{S}+h_{1N}%
\frac{v_{\chi }^{2}}{\Lambda } & 0 \\ 
h_{1N}\frac{v_{\chi }^{2}}{\Lambda } & 0 & h_{1N}\frac{v_{\chi
}^{2}}{\Lambda }%
\end{array}%
\right) =\left( 
\begin{array}{ccc}
A & 0 & A \\ 
0 & B & 0 \\ 
A & 0 & A%
\end{array}%
\right) ,  \notag \\
A &=&-\frac{h_{1N}h_{\rho }^{2}v_{\rho }^{2}v_{\zeta }^{2}v_{\chi
}^{2}v_{\sigma }}{2h_{\chi }^{\left( L\right) }v_{\chi }^{2}\Lambda ^{4}},%
\hspace{1cm}B=-\frac{h_{\rho }^{2}v_{\rho }^{2}v_{\zeta }^{2}v_{\sigma }}{%
2h_{\chi }^{\left( L\right) }v_{\chi }^{2}\Lambda ^{3}}\left( h_{1N}\frac{%
v_{\chi }^{2}}{\Lambda }+\frac{2h_{2N}}{\sqrt{3}}v_{S}+h_{1N}\frac{v_{\chi
}^{2}}{\Lambda }\right) .  \label{Mnu}
\end{eqnarray}

The neutrino mass matrix given in Eq. (\ref{Mnu}) only depends on two
effective parameters: $A$ and $B$. These effective parameters
include the dependence on the various model parameters. It is noteworthy
that $A$ and $B$ are suppressed by inverse powers of the high energy cutoff $\Lambda$ of our model.

\quad The light active neutrino mass matrix $M_{\nu }^{\left( 1\right) }$ is
diagonalized by a unitary rotation matrix $R_{\nu }$, according to: 
\begin{equation}
R_{\nu }^{T}M_{\nu }^{\left( 1\right) }R_{\nu }=\left( 
\begin{array}{ccc}
m_{1} & 0 & 0 \\ 
0 & m_{2} & 0 \\ 
0 & 0 & m_{3}%
\end{array}%
\right) ,\hspace{0.5cm}\mbox{with}\hspace{0.5cm}R_{\nu }=\left( 
\begin{array}{ccc}
\cos \theta & 0 & \sin \theta \\ 
0 & 1 & 0 \\ 
-\sin \theta & 0 & \cos \theta%
\end{array}%
\right) ,\hspace{0.5cm}\theta =\pm \frac{\pi }{4},  \label{Vnu}
\end{equation}%
where the upper sign corresponds to normal ($\theta =+\pi /4$) and the lower
one to inverted ($\theta =-\pi /4$) hierarchy, respectively. The light
active neutrino masses for the normal (NH) and inverted (IH) mass
hierarchies are given by: 
\begin{eqnarray}
\mbox{NH} &:&\theta =+\frac{\pi }{4}:\hspace{10mm}m_{\nu _{1}}=0,\hspace{10mm%
}m_{\nu _{2}}=B,\hspace{10mm}m_{\nu _{3}}=2A,  \label{mass-spectrum-Inverted}
\\[0.12in]
\mbox{IH} &:&\theta =-\frac{\pi }{4}:\hspace{10mm}m_{\nu _{1}}=2A,\hspace{8mm%
}m_{\nu _{2}}=B,\hspace{10mm}m_{\nu _{3}}=0.  \label{mass-spectrum-Normal}
\end{eqnarray}%
We also find that the PMNS leptonic mixing matrix is given by: 
\begin{equation}
U=R_{lL}^{\dag }P_{l}R_{\nu }\simeq \left( 
\begin{array}{ccc}
\frac{\cos \theta }{\sqrt{3}}-\frac{\sin \theta }{\sqrt{3}}e^{i\alpha } & 
\frac{1}{\sqrt{3}} & \frac{\cos \theta }{\sqrt{3}}e^{i\alpha }+\frac{\sin
\theta }{\sqrt{3}} \\ 
&  &  \\ 
\frac{\cos \theta }{\sqrt{3}}-\frac{\sin \theta }{\sqrt{3}}e^{i\alpha +\frac{%
2i\pi }{3}} & \frac{1}{\sqrt{3}}e^{-\frac{2i\pi }{3}} & \frac{\cos \theta }{%
\sqrt{3}}e^{i\alpha +\frac{2i\pi }{3}}+\frac{\sin \theta }{\sqrt{3}} \\ 
&  &  \\ 
\frac{\cos \theta }{\sqrt{3}}-\frac{\sin \theta }{\sqrt{3}}e^{i\alpha -\frac{%
2i\pi }{3}} & \frac{1}{\sqrt{3}}e^{\frac{2i\pi }{3}} & \frac{\cos \theta }{%
\sqrt{3}}e^{i\alpha -\frac{2i\pi }{3}}+\frac{\sin \theta }{\sqrt{3}}%
\end{array}%
\right) .  \label{PMNS}
\end{equation}%
It is worth commenting that the Pontecorvo-Maki-Nakagawa-Sakata (PMNS)
mixing matrix depends only on the parameter $\alpha$, while the neutrino
mass squared splittings are controlled by parameters $A$ and $B$.

The standard parametrization of the leptonic mixing matrix implies that the
lepton mixing angles satisfy \cite{PDG}: 
\begin{eqnarray}
&&\sin ^{2}\theta _{12}=\frac{\left\vert U_{e2}\right\vert ^{2}}{%
1-\left\vert U_{e3}\right\vert ^{2}}=\frac{1}{2-\cos \alpha },\hspace{20mm}%
\sin ^{2}\theta _{13}=\left\vert U_{e3}\right\vert ^{2}=\frac{1}{3}(1+\cos
\alpha ),  \label{theta-ij} \\[3mm]
&&\sin ^{2}\theta _{23}=\frac{\left\vert U_{\mu 3}\right\vert ^{2}}{%
1-\left\vert U_{e3}\right\vert ^{2}}=\frac{1}{2}\left( 1+\frac{\sqrt{3}\sin
\alpha }{\cos \alpha -2}\right) .  \notag
\end{eqnarray}%
The resulting PMNS matrix (\ref{PMNS}) reduces to the trimaximal mixing
matrix (\ref{TBM-ansatz}) in the limit $\alpha =\pi $, for the inverted and
normal hierarchies of the neutrino mass spectrum. Let us note that the
lepton mixing angles are controlled by a single parameter ($\alpha $),
whereas the neutrino mass squared splittings only depend on the parameters $A$ and $B$.

\quad The Jarlskog invariant and the CP violating phase are \cite{PDG}: 
\begin{equation}
J=\func{Im}\left( U_{e1}U_{\mu 2}U_{e2}^{\ast }U_{\mu 1}^{\ast }\right) =-%
\frac{1}{6\sqrt{3}}\cos 2\theta ,\hspace{1cm}\sin \delta =\frac{8J}{\cos
\theta _{13}\sin 2\theta _{12}\sin 2\theta _{23}\sin 2\theta _{13}}.
\end{equation}%
Taking into account that $\theta =\pm \frac{\pi }{4}$, our model predicts $%
J=0$ and $\delta =0$, which results in a vanishing leptonic Dirac CP
violating phase.

\quad In what follows we adjust the three free effective parameters $\alpha $%
, $A$ and $B$ of the lepton sector of our model to reproduce the experimental values of the five
physical observables in the neutrino sector, i.e., three leptonic mixing
parameters and two neutrino mass squared splittings, reported in 
\mbox{Tables
\ref{NH}, \ref{IH}}, for the normal (NH) and inverted (IH) hierarchies of
the neutrino mass spectrum, respectively. We fit only $\alpha $ to adjust
the values of the leptonic mixing parameters $\sin ^{2}\theta _{ij}$,
whereas $A$ and $B$ for the normal (NH) and inverted (IH) mass hierarchies
are given by: 
\begin{equation}
\mbox{NH}: m_{\nu _{1}}=0,\ \ \ m_{\nu _{2}}=B=\sqrt{\Delta m_{21}^{2}}%
\approx 9\mbox{meV},\ \ m_{\nu _{3}}=2A=\sqrt{\Delta m_{31}^{2}}\approx 51%
\mbox{meV};  \label{AB-Delta-IH} 
\end{equation}
\begin{equation}
\mbox{IH}: m_{\nu _{2}}=B=\sqrt{\Delta m_{21}^{2}+\Delta m_{13}^{2}}\approx
50\mbox{meV},\ \ m_{\nu _{1}}=2A=\sqrt{\Delta m_{13}^{2}}\approx 49\mbox{meV}%
,\ m_{\nu _{3}}=0,  \label{AB-Delta-NH}
\end{equation}%
as resulting from Eqs. (\ref{mass-spectrum-Normal}), (\ref%
{mass-spectrum-Inverted}) and the definition $\Delta
m_{ij}^{2}=m_{i}^{2}-m_{j}^{2}$. The best fit values of $\Delta m_{ij}^{2}$
have been taken from Tables \ref{NH} and \ref{IH} for the normal and
inverted mass hierarchies, respectively.

\quad We vary the model parameter $\alpha$ in Eq. (\ref{theta-ij}) to fit
the leptonic mixing parameters $\sin ^{2}\theta _{ij}$ to the experimental
values reported in Tables \ref{NH}, \ref{IH}. We obtain the following best
fit result: 
\begin{eqnarray}
&&\mbox{NH}\ :\ \alpha =-0.88\pi ,\ \ \ \sin ^{2}\theta _{12}\approx 0.34,\
\ \ \sin ^{2}\theta _{23}\approx 0.61,\ \ \ \sin ^{2}\theta _{13}\approx
0.0232;  \label{parameter-fit-IH} \\[0.12in]
&&\mbox{IH}\hspace{2.5mm}:\ \alpha =0.12\,\pi ,\ \ \ \ \ \sin ^{2}\theta
_{12}\approx 0.34,\ \ \ \sin ^{2}\theta _{23}\approx 0.61,\ \ \ \ \,\sin
^{2}\theta _{13}\approx 0.0238.  \label{parameter-fit-NH}
\end{eqnarray}

\quad From the comparison of Eqs. (\ref{parameter-fit-NH}), (\ref%
{parameter-fit-IH}) with Tables \ref{NH}, \ref{IH}, it follows that $\sin
^{2}\theta _{13}$ and $\sin ^{2}\theta _{23}$ are in excellent agreement
with the experimental data, for both normal and inverted mass hierarchies,
whereas $\sin ^{2}\theta _{12}$ is deviated $2\sigma $ away from its best
fit values. This shows that the physical observables in the lepton sector
obtained in our model are consistent with the experimental data.
Furthermore, as previously mentioned, our model predicts a vanishing
leptonic Dirac CP violating phase.

\begin{table}[tbh]
\resizebox{13.5cm}{!}{
\renewcommand{\arraystretch}{1.2}
\begin{tabular}{|c|c|c|c|c|c|}
\hline
Parameter & $\Delta m_{21}^{2}$($10^{-5}$eV$^2$) & $\Delta m_{31}^{2}$($%
10^{-3}$eV$^2$) & $\left( \sin ^{2}\theta _{12}\right) _{\exp }$ & $\left(
\sin ^{2}\theta _{23}\right) _{\exp }$ & $\left( \sin ^{2}\theta
_{13}\right) _{\exp }$ \\ \hline
Best fit & $7.60$ & $2.48$ & $0.323$ & $0.567$ & $0.0234$ \\ \hline
$1\sigma $ range & $7.42-7.79$ & $2.41-2.53$ & $0.307-0.339$ & $0.439-0.599$
& $0.0214-0.0254$ \\ \hline
$2\sigma $ range & $7.26-7.99$ & $2.35-2.59$ & $0.292-0.357$ & $0.413-0.623$
& $0.0195-0.0274$ \\ \hline
$3\sigma $ range & $7.11-8.11$ & $2.30-2.65$ & $0.278-0.375$ & $0.392-0.643$
& $0.0183-0.0297$ \\ \hline
\end{tabular}}%
\caption{Range for experimental values of neutrino mass squared splittings
and leptonic mixing parameters, taken from Ref. \protect\cite{Forero:2014bxa}%
, for the case of normal hierarchy.}
\label{NH}
\end{table}
\begin{table}[tbh]
\resizebox{13.5cm}{!}{
\renewcommand{\arraystretch}{1.2}
\begin{tabular}{|c|c|c|c|c|c|}
\hline
Parameter & $\Delta m_{21}^{2}$($10^{-5}$eV$^{2}$) & $\Delta m_{13}^{2}$($%
10^{-3}$eV$^{2}$) & $\left( \sin ^{2}\theta _{12}\right) _{\exp }$ & $\left(
\sin ^{2}\theta _{23}\right) _{\exp }$ & $\left( \sin ^{2}\theta
_{13}\right) _{\exp }$ \\ \hline
Best fit & $7.60$ & $2.38$ & $0.323$ & $0.573$ & $0.0240$ \\ \hline
$1\sigma $ range & $7.42-7.79$ & $2.32-2.43$ & $0.307-0.339$ & $0.530-0.598$
& $0.0221-0.0259$ \\ \hline
$2\sigma $ range & $7.26-7.99$ & $2.26-2.48$ & $0.292-0.357$ & $0.432-0.621$
& $0.0202-0.0278$ \\ \hline
$3\sigma $ range & $7.11-8.11$ & $2.20-2.54$ & $0.278-0.375$ & $0.403-0.640$
& $0.0183-0.0297$ \\ \hline
\end{tabular}}%
\caption{Range for experimental values of neutrino mass squared splittings
and leptonic mixing parameters, taken from Ref. \protect\cite{Forero:2014bxa}%
, for the case of inverted hierarchy.}
\label{IH}
\end{table}

In the following we proceed to determine the effective Majorana neutrino
mass parameter, which is proportional to the amplitude of neutrinoless
double beta ($0\nu \beta \beta $) decay. This effective Majorana neutrino
mass parameter has the form: 
\begin{equation}
m_{\beta \beta }=\left\vert \sum_{j}U_{ek}^{2}m_{\nu _{k}}\right\vert ,
\label{mee}
\end{equation}%
where $U_{ej}^{2}$ and $m_{\nu _{k}}$ are the PMNS mixing matrix elements
and the Majorana neutrino masses, respectively.

Using Eqs. (\ref{PMNS}), (\ref{AB-Delta-IH}), (\ref{AB-Delta-NH}) and (\ref%
{mee}), it follows that the effective Majorana neutrino mass parameter, for
both Normal and Inverted hierarchies, acquires the following values: 
\begin{equation}
m_{\beta \beta }=\left\{ 
\begin{array}{l}
2\ \mbox{meV}\ \ \ \ \ \ \ \mbox{for \ \ \ \ NH} \\ 
47\ \mbox{meV}\ \ \ \ \ \ \ \mbox{for \ \ \ \ IH} \\ 
\end{array}%
\right.  \label{eff-mass-pred}
\end{equation}%
%
%
%
%
%
%
%
%
%
%
%
%
%
%
%
%
%
%
%
%
%
%
%
%
%
%
%
Our results for the effective Majorana neutrino mass parameter given above,
are beyond the reach of the present and forthcoming $0\nu \beta \beta $
decay experiments. The EXO-200 experiment \cite{Auger:2012ar} sets the
current best upper limit on the effective neutrino mass parameter equal to $%
m_{\beta \beta }\leq 160$ meV, correspoding to $T_{1/2}^{0\nu \beta \beta
}(^{136}\mathrm{Xe})\geq 1.6\times 10^{25}$ yr at 90\% C.L. This bound will
be improved within the not too distant future. The GERDA \textquotedblleft
phase-II\textquotedblright experiment \cite{Abt:2004yk,Ackermann:2012xja} 
is expected to reach 
\mbox{$T^{0\nu\beta\beta}_{1/2}(^{76}{\rm Ge}) \geq
2\times 10^{26}$ yr}, which corresponds to $m_{\beta \beta }\leq 100$ meV. A
bolometric CUORE experiment, using ${}^{130}Te$ \cite{Alessandria:2011rc},
is currently under construction. This experiment features an estimated
sensitivity of about $T_{1/2}^{0\nu \beta \beta }(^{130}\mathrm{Te})\sim
10^{26}$ yr, corresponding to an effective Majorana neutrino mass parameter %
\mbox{$m_{\beta\beta}\leq 50$ meV.} Besides that, there are proposals for
ton-scale next-to-next generation $0\nu \beta \beta $ experiments using $%
^{136}$Xe \cite{KamLANDZen:2012aa,Albert:2014fya} and $^{76}$Ge \cite%
{Abt:2004yk,Guiseppe:2011me}, which claim sensitivities over $T_{1/2}^{0\nu
\beta \beta }\sim 10^{27}$ yr, corresponding to an effective Majorana
neutrino mass parameter $m_{\beta \beta }\sim 12-30$ meV. For a recent
review, see for example Ref. \cite{Bilenky:2014uka}. Consequently, the Eq. (%
\ref{eff-mass-pred}) indicates that our model predicts $T_{1/2}^{0\nu \beta
\beta }$ at the level of sensitivities of the next generation or
next-to-next generation $0\nu \beta \beta $ experiments.

\section{Quark masses and mixing.}

\label{quarkmassesandmixing}From the quark Yukawa terms of Eq. (\ref%
{Lyquarks}) and the relation given by Eq. (\ref{VEVsinglets}), we find that the SM
quarks do not mix with the exotic quarks and that the SM quark mass matrices
are: 
\begin{equation}
M_{U}=\left( 
\begin{array}{ccc}
a_{1}^{\left(U\right) }\lambda ^{8} & 0 & 0 \\ 
0 & a_{2}^{\left(U\right) }\lambda ^{4} & 0 \\ 
0 & 0 & a_{3}^{\left(U\right) }%
\end{array}%
\right) \frac{v}{\sqrt{2}},\hspace{0.7cm}M_{D}=\left( 
\begin{array}{ccc}
a_{11}^{\left(D\right) }\lambda ^{7} & a_{12}^{\left(D\right) }\lambda ^{6}
& a_{13}^{\left(D\right) }\lambda ^{6}e^{-i\delta _{q}} \\ 
a_{21}^{\left(D\right) }\lambda ^{6} & a_{22}^{\left(D\right) }\lambda ^{5}
& a_{23}^{\left(D\right) }\lambda ^{5} \\ 
a_{31}^{\left(D\right) }\lambda ^{6}e^{-i\delta _{q}} & a_{32}^{\left(D\right) }\lambda ^{5} & a_{33}^{\left(D\right) }\lambda ^{3}%
\end{array}%
\right)\frac{v}{\sqrt{2}},
\end{equation}%
where $\lambda =0.225$ is one of the Wolfenstein parameters, $v=246$ GeV the
scale of electroweak symmetry breaking and $a_{ij}^{\left(U,D\right) }$ ($%
i,j=1,2,3$) are $\mathcal{O}(1)$ parameters.

Moreover, we find that the exotic quark masses are: 
\begin{equation}
m_{T}=y^{\left( T\right) }\frac{v_{\chi }}{\sqrt{2}},\hspace{1cm}%
m_{J^{1}}=y_{1}^{\left( J\right) }\frac{v_{\chi }}{\sqrt{2}}=\frac{%
y_{1}^{\left( J\right) }}{y^{\left( T\right) }}m_{T},\hspace{1cm}%
m_{J^{2}}=y_{2}^{\left( J\right) }\frac{v_{\chi }}{\sqrt{2}}=\frac{%
y_{2}^{\left( J\right) }}{y^{\left( T\right) }}m_{T}.  \label{mexotics}
\end{equation}

Since the charged fermion mass and quark mixing pattern emerges from the
breaking of the $Z_{4}\otimes Z_{6}\otimes Z_{16}$ discrete group and in
order to simplify the analysis, the following scenario is considered: 
\begin{equation}
\arg \left( {a_{13}^{\left(D\right) }}\right) =\arg \left( {a_{31}^{\left(D\right) }}\right) ,\hspace{1cm}a_{ij}^{\left(D\right) }=a_{ji}^{\left(D\right) },\hspace{1cm}i,j=1,2,3.
\end{equation}%
Besides that, to show that the quark textures given above can fit the
experimental data, and in order to simplify the analysis, we adopt a
benchmark where we set $a_{1}^{\left(U\right) }=a_{3}^{\left(U\right) }=1$
and $a_{22}^{\left(U\right) }=a_{33}^{\left(D\right) }$, as suggested by
naturalness arguments and by the relation $m_{c}\sim m_{b}$, respectively.
Then, we proceed to fit the parameters $a_{11}^{\left(D\right) }$ $a_{22}^{\left(D\right) }$, $a_{33}^{\left(D\right) }$, $a_{12}^{\left(D\right) }$, $a_{13}^{\left(D\right) }$, $a_{23}^{\left(D\right) }$ and the phase $\delta _{q}$, to reproduce the 10 physical observables of the
quark sector, i.e., the six quark masses, the three mixing angles and the
CP violating phase. The obtained values for the quark masses, the three
quark mixing angles and the CP violating phase $\delta $ in Table \ref{Tab}
correspond to the best fit values: 
\begin{eqnarray}
a_{11}^{\left(D\right) } &\simeq &1.11,\hspace{1cm}a_{22}^{\left(D\right)
}\simeq 0.59,\hspace{1cm}a_{12}^{\left(D\right) }\simeq 0.54,  \notag \\
a_{13}^{\left(D\right) } &\simeq &0.43,\hspace{1cm}a_{23}^{\left(D\right)
}\simeq 1.13,\hspace{1cm}a_{33}^{\left(D\right) }\simeq 1.42,\hspace{1cm}%
\delta _{q}\simeq 66^{\circ }.
\end{eqnarray}
\begin{table}[tbh]
\begin{center}
\begin{tabular}{c|l|l}
\hline\hline
Observable & Model value & Experimental value \\ \hline
$m_{u}(MeV)$ & \quad $1.14$ & \quad $1.45_{-0.45}^{+0.56}$ \\ \hline
$m_{c}(MeV)$ & \quad $635$ & \quad $635\pm 86$ \\ \hline
$m_{t}(GeV)$ & \quad $173.9$ & \quad $172.1\pm 0.6\pm 0.9$ \\ \hline
$m_{d}(MeV)$ & \quad $2.9$ & \quad $2.9_{-0.4}^{+0.5}$ \\ \hline
$m_{s}(MeV)$ & \quad $57.7$ & \quad $57.7_{-15.7}^{+16.8}$ \\ \hline
$m_{b}(GeV)$ & \quad $2.82$ & \quad $2.82_{-0.04}^{+0.09}$ \\ \hline
$\sin \theta _{12}$ & \quad $0.225$ & \quad $0.225$ \\ \hline
$\sin \theta _{23}$ & \quad $0.0412$ & \quad $0.0412$ \\ \hline
$\sin \theta _{13}$ & \quad $0.00352$ & \quad $0.00351$ \\ \hline
$\delta $ & \quad $66^{\circ }$ & \quad $68^{\circ }$ \\ \hline\hline
\end{tabular}%
\end{center}
\caption{Model and experimental values of the quark masses and CKM
parameters.}
\label{Tab}
\end{table}

The obtained quark masses, quark mixing angles and CP violating phase
exhibit an excellent agreement with the experimental data. Let us note, that
despite the aforementioned simplifying assumptions that allow us to
eliminate some of the free parameters, a good fit with the low energy quark
flavor data is obtained, showing that our model is indeed capable of a very
good fit to the experimental data of the physical observables for the quark
sector. The obtained and experimental values for the physical observables of
the quark sector are reported in Table \ref{Tab}. We use the experimental
values of the quark masses at the $M_{Z}$ scale, from Ref. \cite{Bora:2012tx}
(which are similar to those in \cite{Xing:2007fb}), whereas the experimental
values of the CKM parameters are taken from Ref. \cite{PDG}.

In what follows we briefly comment about the phenomenological implications
of our model in the flavor changing processes involving quarks. As
previously mentioned, the different $Z_{3}$ charge assignments for SM and
exotic right handed quark fields imply the absence of mixing between them.
Due to the absence of mixings between SM and exotic quarks, the exotic $T$, $%
J^{1}$ and $J^{2}$ quarks do not exhibit flavor changing neutral decays into
SM quarks and gauge bosons, SM light $126$ GeV Higgs boson and SM quarks.
Thus, assuming that the $H_{2}^{0}$ and $\overline{H}_{2}^{0}$\ neutral
Higgs bosons are heavier than the exotic $T$, $J^{1}$ and $J^{2}$ quarks, it
follows that the flavor changing neutral exotic quark decays are absent in
our model. Consequently these exotic quarks can be searched at the LHC via
their flavor changing charged decays into SM quarks and gauge bosons,
specifically in their dominant decay modes $T\rightarrow bW$ \ and $%
J^{1,2}\rightarrow tW$. These exotic quarks can be produced at the LHC via
Drell-Yan proccesses mediated by charged gauge bosons, where the final
states will include the exotic $T$ quark with a SM down type quark as well
as any of the exotic $J^{1}$ or $J^{2}$ quarks with a SM up type quark.
Furthermore, from the quark Yukawa terms, one can easily see that the our
model predicts the absence of flavor changing top quark decays $t\rightarrow
hc$ and $t\rightarrow hu$ at tree level. The flavor changing top quark
decays $t\rightarrow hc$ and $t\rightarrow hu$\ only arise at one loop level
and will involve virtual charged gauge bosons and exotic quarks running in
the loops. Thus, a measurement of the branching fraction for the $%
t\rightarrow hc$ and $t\rightarrow hu$ decays at the LHC will be crucial for
confirming or ruling out our model. It would be interesting to perform a
detailed study of the exotic quark production at the LHC, the exotic quark
decay modes and the flavor changing top quark decays. This is beyond the
scope of this work and is left for future studies.

\section{Conclusions}

We constructed a predictive $SU(3)_{C}\otimes SU(3)_{L}\otimes U(1)_{X}$
model with $\beta =-\frac{1}{\sqrt{3}}$, \ based on the $A_{4}$ flavor
symmetry supplemented by the $Z_{3}\otimes Z_{4}\otimes Z_{6}\otimes Z_{16}$
discrete group. Our model successfully accounts for the observed fermion
masses and mixing angles. The obtained values for the physical observables
in both quark and lepton sectors exhibit an excellent agreement with the
experimental data. The $A_{4}$, $Z_{4}$ and $Z_{3}$ symmetries allow to
reduce the number of parameters in the Yukawa terms, increasing the
predictivity power of the model. The breaking of the $Z_{4}\otimes
Z_{6}\otimes Z_{16}$ discrete group at high energy, gives rise to the
observed charged fermion mass pattern and quark mixing hierarchy. In our
model the Majorana neutrinos acquire very small masses, much smaller than the
Dirac neutrino masses, thus giving rise to an inverse seesaw mechanism for
the generation of the light active neutrino masses. In this scenario, the
spectrum of neutrinos includes very light active neutrinos and TeV scale
pseudo Dirac nearly degenerate sterile neutrinos. Our model predicts a
vanishing leptonic Dirac CP violating phase as well as an effective Majorana
neutrino mass, relevant for neutrinoless double beta decay, with values $%
m_{\beta \beta }=$ 2 and 48 meV, for the normal and the inverted
hierarchies, respectively. For the inverted hierarchy neutrino mass
spectrum, our obtained value of 48 meV for the effective Majorana neutrino
mass is within the declared reach of the next generation bolometric CUORE
experiment \cite{Alessandria:2011rc} or, more realistically, of the
next-to-next generation tone-scale $0\nu \beta \beta $-decay experiments.
Under the assumption that the exotic $T$, $J^{1}$ and $J^{2}$ quarks are
lighter than the $H_{2}^{0}$ and $\overline{H}_{2}^{0}$\ neutral Higgs
bosons, our model predicts the absence of the flavor changing neutral exotic
quark decays, which implies that they can be searched at the LHC via their
dominant flavor changing charged decay modes $T\rightarrow bW$ \ and $%
J^{1,2}\rightarrow tW$. Furthermore, our model predicts the absence of
flavor changing neutral top quark decays at tree level, implying that they
occur at one loop level. Possible directions for future work along these
lines would be to study the constraints on the heavy charged gauge boson
masses in our model arising from the upper bound on the branching fraction
for the flavor changing top quark decays, the oblique parameters, the $Zb%
\overline{b}$ vertex and the Higgs diphoton signal strength. The heavy
exotic quark decays and their production at the LHC may be useful to study.
All these studies require carefull investigations that we left outside the
scope of this work.

\label{conclusions}

\section*{Acknowledgments}

A.E.C.H was supported by Fondecyt (Chile), Grant No. 11130115 and by DGIP
internal Grant No. 111458. R.M. was supported El Patrimonio Aut\'{o}nomo
Fondo Nacional de Financiamiento para la Ciencia, la Tecnolog\'{\i}a y la
Innovaci\'{o}n Fransisco Jos\'{e} de Caldas programme of COLCIENCIAS in
Colombia.

\appendix

\section{The product rules for $A_4$}

\label{ap1}The $A_{4}$ group has one three-dimensional $\mathbf{3}$\ and
three distinct one-dimensional $\mathbf{1}$, $\mathbf{1}^{\prime }$ and $%
\mathbf{1}^{\prime \prime }$ irreducible representations, satisfying the
following product rules: 
\begin{eqnarray}
&&\hspace{18mm}\mathbf{3}\otimes \mathbf{3}=\mathbf{3}_{s}\oplus \mathbf{3}%
_{a}\oplus \mathbf{1}\oplus \mathbf{1}^{\prime }\oplus \mathbf{1}^{\prime
\prime },  \label{A4-singlet-multiplication} \\[0.12in]
&&\mathbf{1}\otimes \mathbf{1}=\mathbf{1},\hspace{5mm}\mathbf{1}^{\prime
}\otimes \mathbf{1}^{\prime \prime }=\mathbf{1},\hspace{5mm}\mathbf{1}%
^{\prime }\otimes \mathbf{1}^{\prime }=\mathbf{1}^{\prime \prime },\hspace{%
5mm}\mathbf{1}^{\prime \prime }\otimes \mathbf{1}^{\prime \prime }=\mathbf{1}%
^{\prime },  \notag
\end{eqnarray}%
Considering $\left( x_{1},y_{1},z_{1}\right) $ and $\left(
x_{2},y_{2},z_{2}\right) $ as the basis vectors for two $A_{4}$-triplets $%
\mathbf{3}$, the following relations are fullfilled:

\begin{eqnarray}
&&\left( \mathbf{3}\otimes \mathbf{3}\right) _{\mathbf{1}%
}=x_{1}y_{1}+x_{2}y_{2}+x_{3}y_{3},  \label{triplet-vectors} \\
&&\left( \mathbf{3}\otimes \mathbf{3}\right) _{\mathbf{3}_{s}}=\left(
x_{2}y_{3}+x_{3}y_{2},x_{3}y_{1}+x_{1}y_{3},x_{1}y_{2}+x_{2}y_{1}\right) ,\
\ \ \ \left( \mathbf{3}\otimes \mathbf{3}\right) _{\mathbf{1}^{\prime
}}=x_{1}y_{1}+\omega x_{2}y_{2}+\omega ^{2}x_{3}y_{3},  \notag \\
&&\left( \mathbf{3}\otimes \mathbf{3}\right) _{\mathbf{3}_{a}}=\left(
x_{2}y_{3}-x_{3}y_{2},x_{3}y_{1}-x_{1}y_{3},x_{1}y_{2}-x_{2}y_{1}\right) ,\
\ \ \left( \mathbf{3}\otimes \mathbf{3}\right) _{\mathbf{1}^{\prime \prime
}}=x_{1}y_{1}+\omega ^{2}x_{2}y_{2}+\omega x_{3}y_{3},  \notag
\end{eqnarray}%
where $\omega =e^{i\frac{2\pi }{3}}$. The representation $\mathbf{1}$ is
trivial, while the non-trivial $\mathbf{1}^{\prime }$ and $\mathbf{1}%
^{\prime \prime }$ are complex conjugate to each other. Some reviews of
discrete symmetries in particle physics are found in Refs. \cite%
{King:2013eh,Altarelli:2010gt,Ishimori:2010au,Discret-Group-Review}.


\begin{thebibliography}{()}
\section*{References}
\bibitem[()]{atlashiggs} G.~Aad \textit{et al.} [The ATLAS Collaboration],
``Observation of a new particle in the search for the Standard Model Higgs
boson with the ATLAS detector at the LHC,'', Phys. Lett. B \textbf{716}
(2012) 1 [\href{http://arxiv.org/abs/hep-ex/1207.7214}{arXiv:hep-ex/1207.7214%
}]. 

\bibitem[()]{cmshiggs} S.~Chatrchyan \textit{et al.} [The CMS
Collaboration], ``Observation of a new boson at a mass of 125 GeV with the
CMS experiment at the LHC,'' , Phys. Lett. B \textbf{716}, 30 (2012) [\href{http://arxiv.org/abs/hep-ex/1207.7235}%
{arXiv:hep-ex/1207.7235}]. 

\bibitem[()]{newtevatron} T.~Aaltonen \textit{et al.} [CDF and D0
Collaborations], ``Evidence for a particle produced in association with weak
bosons and decaying to a bottom-antibottom quark pair in Higgs boson
searches at the Tevatron,'', Phys. Rev. Lett. \textbf{109} (2012) 071804, [ 
\href{http://arxiv.org/abs/hep-ex/1207.6436}{arXiv:hep-ex/1207.6436}]. 


\bibitem[()]{CMS-PAS-HIG-12-020} The CMS Collaboration, ``Observation of a
new boson with a mass near 125 GeV,'' CMS-PAS-HIG-12-020.

\bibitem[()]{SM} S.L. Glashow, Nucl. Phys. 22, 579 (1961); S. Weinberg,
Phys. Rev. Lett. \textbf{19}, 1264 (1967); A. Salam, in \textit{Elementary
Particle Theory: Relativistic Groups and Analyticity (Nobel Symposium No. 8)}%
, edited by N.Svartholm (Almqvist and Wiksell, Stockholm, 1968), p. 367.

\bibitem[()]{PDG} 
K.~A.~Olive \textit{et al.} [Particle Data Group Collaboration], 
Chin.\ Phys.\ C \textbf{38}, 090001 (2014).
doi:10.1088/1674-1137/38/9/090001 

\bibitem[()]{BSMtheorieswithDM} J.~K.~Mizukoshi, C.~A.~de S.Pires,
F.~S.~Queiroz and P.~S.~Rodrigues da Silva, 
Phys.\ Rev.\ \textbf{D 83}, 065024 (2011) [arXiv:1010.4097 [hep-ph]];
J.~D.~Ruiz-Alvarez, C.~A.~de S.Pires, F.~S.~Queiroz, D.~Restrepo and
P.~S.~Rodrigues da Silva, 
Phys.\ Rev.\ D \textbf{86}, 075011 (2012) doi:10.1103/PhysRevD.86.075011
[arXiv:1206.5779 [hep-ph]]; C.~Kelso, C.~A.~de S. Pires, S.~Profumo,
F.~S.~Queiroz and P.~S.~Rodrigues da Silva, 
Eur.\ Phys.\ J.\ C \textbf{74}, 2797 (2014) [arXiv:1308.6630 [hep-ph]];
S.~Profumo and F.~S.~Queiroz, 
Eur.\ Phys.\ J.\ C \textbf{74}, 2960 (2014) [arXiv:1307.7802 [hep-ph]];
C.~Kelso, H.~N.~Long, R.~Martinez and F.~S.~Queiroz, 
Phys.\ Rev.\ D \textbf{90}, 113011 (2014) [arXiv:1408.6203 [hep-ph]];
R.~Martinez, J.~Nisperuza, F.~Ochoa and J.~P.~Rubio, 
Phys.\ Rev.\ D \textbf{90}, 095004 (2014) [arXiv:1408.5153 [hep-ph]];
H.~Okada and Y.~Orikasa, 
Phys.\ Rev.\ D \textbf{90}, no. 7, 075023 (2014) [arXiv:1407.2543 [hep-ph]];
D.~Cogollo, A.~X.~Gonzalez-Morales, F.~S.~Queiroz and P.~R.~Teles, 
JCAP \textbf{1411}, no. 11, 002 (2014) doi:10.1088/1475-7516/2014/11/002
[arXiv:1402.3271 [hep-ph]]; H.~Hatanaka, K.~Nishiwaki, H.~Okada and
Y.~Orikasa, 
arXiv:1412.8664 [hep-ph]; P.~V.~Dong, C.~S.~Kim, D.~V.~Soa and N.~T.~Thuy, 
arXiv:1501.04385 [hep-ph]; R.~Martinez and F.~Ochoa, 
arXiv:1512.04128 [hep-ph]; I.~Medeiros Varzielas and O.~Fischer,  
JHEP \textbf{1601}, 160 (2016)  doi:10.1007/JHEP01(2016)160 
[arXiv:1512.00869 [hep-ph]]. 


\bibitem{King:2013eh} S.~F.~King and C.~Luhn, 
arXiv:1301.1340 [hep-ph]. 


\bibitem{Altarelli:2010gt} G.~Altarelli and F.~Feruglio, 
Rev.\ Mod.\ Phys.\ \textbf{82} (2010) 2701 [arXiv:1002.0211 [hep-ph]]. 


\bibitem{Ishimori:2010au} H.~Ishimori, T.~Kobayashi, H.~Ohki, Y.~Shimizu,
H.~Okada and M.~Tanimoto, 
Prog.\ Theor.\ Phys.\ Suppl.\ \textbf{183} (2010) 1 [arXiv:1003.3552
[hep-th]]. 


\bibitem[()]{StringsandDS} 
T.~Kobayashi, H.~P.~Nilles, F.~Ploger, S.~Raby and M.~Ratz, 
Nucl.\ Phys.\ B \textbf{768}, 135 (2007) [hep-ph/0611020]; 
T.~Kobayashi, Y.~Omura and K.~Yoshioka, 
Phys.\ Rev.\ D \textbf{78}, 115006 (2008) [arXiv:0809.3064 [hep-ph]]; 
H.~Abe, K.~S.~Choi, T.~Kobayashi and H.~Ohki, 
Nucl.\ Phys.\ B \textbf{820}, 317 (2009) [arXiv:0904.2631 [hep-ph]]; 
M.~Berasaluce-Gonzalez, L.~E.~Ibanez, P.~Soler and A.~M.~Uranga, 
JHEP \textbf{1112}, 113 (2011) [arXiv:1106.4169 [hep-th]]; 
F.~Beye, T.~Kobayashi and S.~Kuwakino, 
Phys.\ Lett.\ B \textbf{736}, 433 (2014) [arXiv:1406.4660 [hep-th]]. 


\bibitem[()]{331-pisano} F. Pisano and V. Pleitez, Phys. Rev. \textbf{D46},
410 (1992); Nguyen Tuan Anh, Nguyen Anh Ky, Hoang Ngoc Long, Int. J. Mod.
Phys. \textbf{A16}, 541 (2001).

\bibitem[()]{331-frampton} P.H. Frampton, Phys. Rev. Lett. \textbf{69}, 2889
(1992).

\bibitem[()]{331-long} R. Foot, H.N. Long and T.A. Tran, Phys. Rev. \textbf{%
D50}, R34 (1994); H.N. Long, ibid. \textbf{53}, 437 (1996); \textbf{54},
4691 (1996); Mod. Phys. Lett. \textbf{A 13}, 1865 (1998).

\bibitem[()]{M-O} R.~A.~Diaz, R.~Martinez, J.~Mira and J.~A.~Rodriguez, 
Phys.\ Lett.\ B \textbf{552}, 287 (2003) [hep-ph/0208176]; Rodolfo A. Diaz,
R. Martinez, F. Ochoa, Phys. Rev. \textbf{D69}, 095009 (2004); \textbf{D72},
035018 (2005); Fredy Ochoa, R. Martinez, Phys. Rev \textbf{D72}, 035010
(2005); A. Carcamo, R. Martinez and F. Ochoa, Phys. Rev. \textbf{D73},
035007 (2006); 
C. Alvarado, R. Mart\'{\i}nez and F. Ochoa, Phys. Rev. \textbf{D86}, 025027
(2012) 
A.~E.~C\'{a}rcamo Hern\'{a}ndez, R.~Mart\'{\i}nez and F.~Ochoa, 
Phys.\ Rev.\ \textbf{D 87} (2013) 075009 [arXiv:1302.1757 [hep-ph]];
P.~V.~Dong, T.~P.~Nguyen and D.~V.~Soa, 
Phys.\ Rev.\ D \textbf{88}, no. 9, 095014 (2013)
doi:10.1103/PhysRevD.88.095014 [arXiv:1308.4097 [hep-ph]]; 
J.~W.~F.~Valle and C.~A.~Vaquera-Araujo, 
arXiv:1601.05237 [hep-ph];

\bibitem[()]{anomalias} J.S. Bell, R. Jackiw, Nuovo Cim. \textbf{A60}, 47
(1969); S.L. Adler, Phys. Rev. \textbf{177}, 2426 (1969); D.J. Gross, R.
Jackiw, Phys.Rev. \textbf{D6}, 477 (1972). H. Georgi and S. L. Glashow,
Phys. Rev. \textbf{D6}, 429 (1972); S. Okubo, Phys. Rev.\textbf{\ D16}, 3528
(1977); J. Banks and H. Georgi, Phys. Rev. \textbf{14}, 1159 (1976).

\bibitem{diphotonexcess331}A.~E.~C.~Hern\'andez and I.~Nisandzic, 
arXiv:1512.07165 [hep-ph]; S.~M.~Boucenna, S.~Morisi and A.~Vicente, 
arXiv:1512.06878 [hep-ph]; Q.~H.~Cao, Y.~Liu, K.~P.~Xie, B.~Yan and
D.~M.~Zhang, 
arXiv:1512.08441 [hep-ph];  P.~V.~Dong and N.~T.~K.~Ngan,
  arXiv:1512.09073 [hep-ph].

\bibitem{dibosonexcess331}Q.~H.~Cao, B.~Yan
and D.~M.~Zhang, 
Phys.\ Rev.\ D \textbf{92}, no. 9, 095025 (2015)
doi:10.1103/PhysRevD.92.095025 [arXiv:1507.00268 [hep-ph]]


\bibitem[()]{An:2012eh} F.~P.~An \textit{et al.} (DAYA-BAY Collaboration), 
Phys.\ Rev.\ Lett.\ \textbf{108}, 171803 (2012).

\bibitem[()]{Abe:2011sj} K.~Abe \textit{et al.} (T2K Collaboration), 
Phys.\ Rev.\ Lett.\ \textbf{107}, 041801 (2011).%

\bibitem[()]{Adamson:2011qu} P.~Adamson \textit{et al.} (MINOS
Collaboration), 
Phys.\ Rev.\ Lett.\ \textbf{107}, 181802 (2011).%

\bibitem[()]{Abe:2011fz} Y.~Abe \textit{et al.} (DOUBLE-CHOOZ
Collaboration), 
Phys.\ Rev.\ Lett.\ \textbf{108}, 131801 (2012).%

\bibitem[()]{Ahn:2012nd} J.~K.~Ahn \textit{et al.} (RENO Collaboration), 
Phys.\ Rev.\ Lett.\ \textbf{108}, 191802 (2012).%
















\bibitem{Ade:2013zuv} P.~A.~R.~Ade \textit{et al.} [Planck Collaboration], 
Astron.\ Astrophys.\ \textbf{571}, A16 (2014) [arXiv:1303.5076
[astro-ph.CO]].



\bibitem[()]{Kraus:2004zw} C.~Kraus, B.~Bornschein, L.~Bornschein, J.~Bonn,
B.~Flatt, A.~Kovalik, B.~Ostrick and E.~W.~Otten \textit{et al.}, 
Eur.\ Phys.\ J.\ C \textbf{40}, 447 (2005) [hep-ex/0412056]. 


\bibitem[()]{Auger:2012ar} EXO Collaboration, M.~Auger \emph{et~al.}, %
\newblock Phys.Rev.Lett. \textbf{109}, 032505 (2012), arXiv:1205.5608. 

\bibitem[()]{Abt:2004yk} GERDA Collaboration, I.~Abt \emph{et~al.}, \newblock%
(2004), arXiv:hep-ex/0404039. 

\bibitem[()]{Ackermann:2012xja} GERDA Collaboration, K.-H. Ackermann \emph{%
et~al.}, \newblock (2012), arXiv:1212.4067. 

\bibitem[()]{Alessandria:2011rc} F.~Alessandria \emph{et~al.}, \newblock %
(2011),arXiv:1109.0494. 

\bibitem[()]{KamLANDZen:2012aa} KamLAND-Zen Collaboration, A.~Gando \emph{%
et~al.} 

\bibitem[()]{Auty:2013:zz} EXO-200 Collaboration, D.~Auty, \newblock %
Recontres de Moriond, http://moriond.in2p3.fr/ (2013).

\bibitem[()]{Guiseppe:2011me} Majorana Collaboration, C.~Aalseth \emph{et~al.%
}, \newblock Nucl.Phys.Proc.Suppl. \textbf{217}, 44 (2011), arXiv:1101.0119. 


\bibitem[()]{Albert:2014fya} J.~B.~Albert \textit{et al.} [EXO-200
Collaboration], 
Phys.\ Rev.\ D \textbf{90}, no. 9, 092004 (2014) [arXiv:1409.6829 [hep-ex]]. 


\bibitem[()]{Bilenky:2014uka} 
O.~Cremonesi and M.~Pavan, 
arXiv:1310.4692 [physics.ins-det]; W.~Rodejohann, 
J.\ Phys.\ G \textbf{39}, 124008 (2012) [arXiv:1206.2560 [hep-ph]];
A.~Barabash, \newblock (2012), arXiv:1209.4241; F.~F.~Deppisch, M.~Hirsch
and H.~Pas, 
J.\ Phys.\ G \textbf{39}, 124007 (2012) [arXiv:1208.0727 [hep-ph]];
A.~Giuliani and A.~Poves, 
Adv.\ High Energy Phys.\ \textbf{2012}, 857016 (2012); 
S.~M.~Bilenky and C.~Giunti, 
arXiv:1411.4791 [hep-ph]. 


\bibitem[()]{Forero:2014bxa} D.~V.~Forero, M.~Tortola and J.~W.~F.~Valle, 
Phys.\ Rev.\ D \textbf{90}, no. 9, 093006 (2014) [arXiv:1405.7540 [hep-ph]].


\bibitem[()]{discrete-lepton} E. Ma and G. Rajasekaran, Phys. Rev. \textbf{D
64}, 113012 (2001); G. Altarelli and F. Feruglio, Nucl. Phys. \textbf{B 741}%
, 215 (2006); S. L. Chen, M. Frigerio and E. Ma, Nucl. Phys. \textbf{B 724},
423 (2005); A. Zee, Phys. Lett. \textbf{B 630}, 58 (2005); C.~Csaki,
C.~Delaunay, C.~Grojean and Y.~Grossman, 
JHEP \textbf{0810}, 055 (2008) doi:10.1088/1126-6708/2008/10/055
[arXiv:0806.0356 [hep-ph]]; M.~Hirsch, S.~Morisi and J.~W.~F.~Valle, 
Phys.\ Lett.\ B \textbf{679}, 454 (2009) [arXiv:0905.3056 [hep-ph]];
P.~V.~Dong, H.~N.~Long, D.~V.~Soa and V.~V.~Vien, 
Eur.\ Phys.\ J.\ C \textbf{71}, 1544 (2011)
doi:10.1140/epjc/s10052-011-1544-2 [arXiv:1009.2328 [hep-ph]]; 
T~Fukuyama, H.~Sugiyama and K.~Tsumura, 
Phys.\ Rev.\ \textbf{D} \textbf{83}, 056016 (2011) [arXiv:1012.4886
[hep-ph]]; 
I.~de Medeiros Varzielas and Lu\'{\i}s~Lavoura, 
J.\ Phys.\ G \textbf{40}, 085002 (2013) [arXiv:1212.3247 [hep-ph]];
P.~M.~Ferreira, L.~Lavoura and P.~O.~Ludl, 
arXiv:1306.1500 [hep-ph]; 
P.~V.~Dong, L.~T.~Hue, H.~N.~Long and D.~V.~Soa, 
Phys.\ Rev.\ D \textbf{81}, 053004 (2010) [arXiv:1001.4625 [hep-ph]];
P.~S.~Bhupal Dev, B.~Dutta, R.~N.~Mohapatra and M.~Severson, 
Phys.\ Rev.\ D \textbf{86}, 035002 (2012) doi:10.1103/PhysRevD.86.035002
[arXiv:1202.4012 [hep-ph]]; V.~V.~Vien and H.~N.~Long, 
Int.\ J.\ Mod.\ Phys.\ A \textbf{28}, 1350159 (2013)
doi:10.1142/S0217751X13501595 [arXiv:1312.5034 [hep-ph]]; 
A.~C.~B.~Machado, J.~C.~Montero and V.~Pleitez, 
Phys.\ Lett.\ B \textbf{697}, 318 (2011) [arXiv:1011.5855 [hep-ph]]; Miguel
D. Campos, A.~E.~C\'{a}rcamo H\'{e}rnandez, H.~Pas and Erik Schumacher,
arXiv:1408.1652 [hep-ph]; A.~E.~C\'{a}rcamo Hern\'{a}ndez, S.~G.~Kovalenko
and I.~Schmidt, arXiv:1411.2913 [hep-ph]; A.~E.~C\'{a}rcamo Hern\'{a}ndez,
R. Martinez and F.~Ochoa, 
arXiv:1309.6567 [hep-ph]; A.~E.~C\'{a}rcamo Hern\'{a}ndez, R. Martinez and
Jorge Nisperuza, 
Eur.\ Phys.\ J.\ C \textbf{75}, no. 2, 72 (2015) [arXiv:1401.0937 [hep-ph]];
I.~de Medeiros Varzielas, O.~Fischer and V.~Maurer, 
JHEP \textbf{1508}, 080 (2015) doi:10.1007/JHEP08(2015)080 [arXiv:1504.03955
[hep-ph]]; V.~V.~Vien, H.~N.~Long and D.~P.~Khoi, 
Int.\ J.\ Mod.\ Phys.\ A \textbf{30}, no. 17, 1550102 (2015)
doi:10.1142/S0217751X1550102X [arXiv:1506.06063 [hep-ph]]; B.~Karmakar and
A.~Sil, 
Phys.\ Rev.\ D \textbf{91}, 013004 (2015) doi:10.1103/PhysRevD.91.013004
[arXiv:1407.5826 [hep-ph]]; A.~Dev, P.~Ramadevi and S.~U.~Sankar, 
JHEP \textbf{1511}, 034 (2015) doi:10.1007/JHEP11(2015)034 [arXiv:1504.04034
[hep-ph]]; A.~E.~C\'{a}rcamo Hern\'{a}ndez, I.~de Medeiros Varzielas and
E.~Schumacher, 
arXiv:1509.02083 [hep-ph]; A.~E.~C\'{a}rcamo Hern\'{a}ndez, I.~de Medeiros
Varzielas and Nicol\'as~A.~Neill, arXiv:1511.07420 [hep-ph]; A.~E.~C\'{a}%
rcamo Hern\'{a}ndez, I.~d.~M.~Varzielas and E.~Schumacher, 
arXiv:1601.00661 [hep-ph]; C.~Arbela\'ez, A.~E.~C\'{a}rcamo Hern\'{a}ndez,
S.~Kovalenko and I.~Schmidt, 
arXiv:1602.03607 [hep-ph].

\bibitem[()]{6HDMA4} A.~E.~C\'{a}rcamo Hern\'{a}ndez, I.~d.~M.~Varzielas,
S.~G.~Kovalenko, H.~Pas and I.~Schmidt, Phys.\ Rev.\ D \textbf{88} 076014
(2013) [arXiv:1307.6499 [hep-ph]]. 

\bibitem[()]{SU5A4} Miguel D. Campos, A.~E.~C\'{a}rcamo Hern\'{a}ndez,
S.~G.~Kovalenko, I.~Schmidt and Erik Schumacher, Phys.\ Rev.\ D \textbf{90}
016006 (2014) [arXiv:1403.2525 [hep-ph]].

\bibitem[()]{discrete-quark} F. Feruglio, C. Hagedorn, Y. Lin and L. Merlo,
Nucl. Phys. B \textbf{775}, 120 (2007); M. C. Chen and K. T. Mahanthappa,
Phys. Lett. B \textbf{652}, 34 (2007); S.~Morisi, M.~Nebot, K.~M.~Patel,
E.~Peinado and J.~W.~F.~Valle, 
arXiv:1303.4394 [hep-ph]; Z.~z.~Xing, D.~Yang, S.~Zhou,
[arXiv:hep-ph/1004.4234v2]; 
J.~E.~Kim, M.~S.~Seo, JHEP \textbf{1102} (2011) 097,
[arXiv:hep-ph/1005.4684]; 
I.~de Medeiros Varzielas and D.~Pidt, 
arXiv:1307.0711 [hep-ph]. 

\bibitem[()]{s3pheno} S.~Pakvasa and H.~Sugawara, 
Phys.\ Lett.\ B \textbf{73} (1978) 61; 
E.~Ma, 
Phys.\ Rev.\ D \textbf{61}, 033012 (2000) [arXiv:hep-ph/9909249]; 
W.~Grimus and L.~Lavoura, 
JHEP \textbf{0508}, 013 (2005) [arXiv:hep-ph/0504153]; G.~Bhattacharyya,
P.~Leser and H.~Pas, 
Phys.\ Rev.\ D \textbf{83}, 011701 (2011) 
[arXiv:1006.5597 [hep-ph]]; G.~Bhattacharyya, P.~Leser and H.~Pas, 
Phys.\ Rev.\ D \textbf{86}, 036009 (2012) 
[arXiv:1206.4202 [hep-ph]]; P.~V.~Dong, H.~N.~Long, C.~H.~Nam and
V.~V.~Vien, Phys.\ Rev.\ D \textbf{85}, 053001 (2012) [arXiv:1111.6360
[hep-ph]]; F.~Gonz\'{a}lez~Canales, A.~Mondrag\'{o}n, M.~Mondrag\'{o}n,
U.~J.~Saldana~Salazar and L.~Velasco-Sevilla, 
Phys.\ Rev.\ D \textbf{88}, 096004 (2013) [arXiv:1304.6644 [hep-ph]];
Y.~Kajiyama, H.~Okada and K.~Yagyu, arXiv:1309.6234 [hep-ph]. 

\bibitem[()]{331S3} A.~E.~C\'arcamo Hern\'andez, M.E. Cata\~{n}o and
R.~Martinez, 
Phys.\ Rev.\ D \textbf{90}, 073001 (2014) [arXiv:1407.5217 [hep-ph]].

\bibitem[()]{Delta27} E.~Ma, 
Mod.\ Phys.\ Lett.\ A \textbf{21}, 1917 (2006) 
[hep-ph/0607056]; I.~d.~M.~Varzielas and D.~Pidt, 
arXiv:1307.0711 [hep-ph]; G.~Bhattacharyya, I.M. Varzielas and P.~Leser, 
Phys.\ Rev.\ Lett.\ \textbf{109}, 241603 (2012) [arXiv:1210.0545 [hep-ph]];
C.~C.~Nishi, 
arXiv:1306.0877 [hep-ph]; I.~de Medeiros Varzielas, 
JHEP \textbf{1508}, 157 (2015) doi:10.1007/JHEP08(2015)157 [arXiv:1507.00338
[hep-ph]]; G.~C.~Branco, I.~de Medeiros Varzielas and S.~F.~King, 
Nucl.\ Phys.\ B \textbf{899}, 14 (2015) doi:10.1016/j.nuclphysb.2015.07.024
[arXiv:1505.06165 [hep-ph]]; 
P.~Chen, G.~J.~Ding, A.~D.~Rojas, C.~A.~Vaquera-Araujo and J.~W.~F.~Valle, 
arXiv:1509.06683 [hep-ph]; V.~V.~Vien, A.~E.~C\'arcamo Hern\'andez and
H.~N.~Long, 
arXiv:1601.03300 [hep-ph]; 
A.~E.~C\'arcamo Hern\'andez, H.~N.~Long and V.~V.~Vien, 
arXiv:1601.05062 [hep-ph]. 

\bibitem[()]{T7} C.~Luhn, S.~Nasri and P.~Ramond, 
Phys.\ Lett.\ B \textbf{652}, 27 (2007) [arXiv:0706.2341 [hep-ph]];
C.~Hagedorn, M.~A.~Schmidt and A.~Y.~Smirnov, 
Phys.\ Rev.\ D \textbf{79}, 036002 (2009) [arXiv:0811.2955 [hep-ph]];
Q.~H.~Cao, S.~Khalil, E.~Ma and H.~Okada, 
Phys.\ Rev.\ Lett.\ \textbf{106}, 131801 (2011) 
[arXiv:1009.5415 [hep-ph]]; C.~Luhn, K.~M.~Parattu and A.~Wingerter, 
JHEP \textbf{1212}, 096 (2012) [arXiv:1210.1197 [hep-ph]]; Y.~Kajiyama,
H.~Okada and K.~Yagyu, 
JHEP \textbf{1310}, 196 (2013) 
[arXiv:1307.0480 [hep-ph]]; H.~Ishimori, S.~Khalil and E.~Ma, 
Phys.\ Rev.\ D \textbf{86}, 013008 (2012) [arXiv:1204.2705 [hep-ph]]; 
V.~V.~Vien and H.~N.~Long, 
JHEP \textbf{1404}, 133 (2014) [arXiv:1402.1256 [hep-ph]]; C.~Bonilla,
S.~Morisi, E.~Peinado and J.~W.~F.~Valle, 
arXiv:1411.4883 [hep-ph]; A.~E.~C\'{a}rcamo Hern\'{a}ndez and and
R.~Martinez, 
arXiv:1501.07261 [hep-ph]; A.~E.~C\'{a}rcamo Hern\'{a}ndez and R.~Martinez, 
arXiv:1511.07997 [hep-ph].

\bibitem[()]{SU5T7} Carolina Arbela\'ez, A.~E.~C\'{a}rcamo Hern\'{a}ndez,
Sergey Kovalenko and Iv\'an Schmidt, 
Phys.\ Rev.\ D \textbf{92}, no. 11, 115015 (2015)
doi:10.1103/PhysRevD.92.115015 [arXiv:1507.03852 [hep-ph]].

\bibitem[()]{Tprime} P.~H.~Frampton, T.~W.~Kephart and S.~Matsuzaki, 
Phys.\ Rev.\ D \textbf{78}, 073004 (2008) [arXiv:0807.4713 [hep-ph]];
P.~H.~Frampton, C.~M.~Ho and T.~W.~Kephart, 
Phys.\ Rev.\ D \textbf{89}, 027701 (2014) [arXiv:1305.4402 [hep-ph]].

\bibitem[()]{textures} H. Fritzsch, Phys. Lett. \textbf{B70}, 436 (1977); H.
Fritzsch, Phys. Lett. \textbf{B73}, 317 (1978); H. Fritzsch, Nucl. Phys. 
\textbf{B155}, 189 (1979); T.P. Cheng and M. Sher, Phys. Rev. \textbf{D35},
3484 (1987); H.~Fritzsch and J.~Planck, Phys.\ Lett.\ B \textbf{237}, 451
(1990); 
D.~s.~Du and Z.~z.~Xing, 
Phys.\ Rev.\ D \textbf{48}, 2349 (1993); R.~G.~Roberts, A.~Romanino,
G.~G.~Ross and L.~Velasco-Sevilla, 
Nucl.\ Phys.\ B \textbf{615}, 358 (2001) 
[hep-ph/0104088]; T.~Kikuchi and T.~Fukuyama, 
Phys.\ Rev.\ D \textbf{65}, 097301 (2002) 
[hep-ph/0202189]; 
K. Matsuda and H. Nishiura, Phys. Rev. \textbf{D74}, 033014 (2006); A.~E.~C%
\'{a}rcamo Hern\'{a}ndez, R.~Martinez and J.~A.~Rodriguez, Eur. Phys. J. 
\textbf{C50}, 935 (2007); A.~E.~C\'{a}rcamo Hern\'{a}ndez, R.~Martinez and
J.~A.~Rodriguez, AIP Conf.\ Proc.\ \textbf{1026} (2008) 272; H. Okada and K.
Yagyu, Phys.\ Rev.\ D \textbf{89} 053008 (2014) [arXiv:1311.4360 [hep-ph]];
H. Okada and K. Yagyu, arXiv:1405.2368 [hep-ph]; H.~Pas and E.~Schumacher, 
Phys.\ Rev.\ D \textbf{89}, no. 9, 096010 (2014) 
[arXiv:1401.2328 [hep-ph]]; A. E. C\'{a}rcamo Hern\'{a}ndez and
I.~d.~M.~Varzielas, 
J.\ Phys.\ G \textbf{42}, no. 6, 065002 (2015)
doi:10.1088/0954-3899/42/6/065002 [arXiv:1410.2481 [hep-ph]]; A.~Palcu, 
arXiv:1408.6518 [hep-ph]; 
R.~Sinha, R.~Samanta and A.~Ghosal, 
arXiv:1508.05227 [hep-ph]; H.~Nishiura and T.~Fukuyama, 
arXiv:1510.01035 [hep-ph]; R.~R.~Gautam, M.~Singh and M.~Gupta, 
Phys.\ Rev.\ D \textbf{92}, no. 1, 013006 (2015) 
[arXiv:1506.04868 [hep-ph]]; H.~Okada and Y.~Orikasa, 
arXiv:1509.04068 [hep-ph]; 
H.~Pas and E.~Schumacher, 
arXiv:1510.08757 [hep-ph]; A.~Palcu, 
arXiv:1510.06717 [hep-ph]; A. E. C\'{a}rcamo Hern\'{a}ndez, 
arXiv:1512.09092 [hep-ph]. 

\bibitem[()]{GUT} R.~Barbieri, G.~R.~Dvali, A.~Strumia, Z.~Berezhiani and
L.~J.~Hall, 
Nucl.\ Phys.\ B \textbf{432}, 49 (1994) [arXiv:hep-ph/9405428]; 
Z.~Berezhiani, 
Phys.\ Lett.\ B \textbf{355}, 481 (1995) [arXiv:hep-ph/9503366]; 
A.~E.~C\'arcamo Hern\'andez and Rakibur Rahman [arXiv:hep-ph/1007.0447]. 

\bibitem[()]{Extradim} B.~A.~Dobrescu, 
Phys.\ Lett.\ B \textbf{461}, 99 (1999) [arXiv:hep-ph/9812349]; 
G.~Altarelli and F.~Feruglio, 
Nucl.\ Phys.\ B \textbf{720} 64 (2005) [hep-ph/0504165]; 
A.~E.~C\'arcamo Hern\'andez, Claudio.~O.~Dib, Nicol\'as~Neill H and
Alfonso~R.~Zerwekh, JHEP \textbf{1202} (2012) 132 [arXiv:hep-ph/1201.0878];
M.~Frank, C.~Hamzaoui, N.~Pourtolami and M.~Toharia, 
Phys.\ Lett.\ B \textbf{742}, 178 (2015) 
[arXiv:1406.2331 [hep-ph]]; A. E. C\'{a}rcamo Hern\'{a}ndez,
I.~d.~M.~Varzielas and N.~A.~Neill, 
arXiv:1511.07420 [hep-ph]. 

\bibitem[()]{String} K.~S.~Babu and R.~N.~Mohapatra, 
Phys.\ Rev.\ Lett.\ \textbf{74}, 2418 (1995) [arXiv:hep-ph/9410326].

\bibitem[()]{horizontal} L. E. Ibanez and G. G. Ross, Phys. Lett. \textbf{B
332}, 100 (1994); P. Binetruy and P. Ramond, Phys. Lett. \textbf{B 350}, 49
(1995); Y. Nir, Phys. Lett. \textbf{B 354}, 107 (1995); V. Jain and R.
Shrock, Phys. Lett.\textbf{\ B 352}, 83 (1995); E. Dudas, S. Pokorski and C.
A. Savoy, Phys. Lett. \textbf{B 356}, 45 (1995); \textbf{B 369}, 255 (1996).


%



\bibitem[()]{catano} 
M.E. Catano, R. Martinez and F. Ochoa, 
Phys.\ Rev.\ D \textbf{86}, 073015 (2012) [arXiv:1206.1966 [hep-ph]].

\bibitem[()]{grimus} W. Grimus and L. Lavoura, JHEP \textbf{0011}, 042
(2000).


\bibitem[()]{Bora:2012tx} K.~Bora, 
J.\ Phys.\ \textbf{2}, 2013 [arXiv:1206.5909 [hep-ph]].

\bibitem[()]{Xing:2007fb} Z.~z.~Xing, H.~Zhang and S.~Zhou, 
Phys.\ Rev.\ D \textbf{77}, 113016 (2008) [arXiv:0712.1419 [hep-ph]]. 




\bibitem[()]{Discret-Group-Review} P. Ramond, \textit{Group Theory: A
Physicist's Survey}, Cambridge University Press, UK (2010).

\end{thebibliography}
\end{document}